\documentclass[a4paper,10pt]{article}
\usepackage[utf8]{inputenc}

\usepackage{graphicx}
\usepackage{epstopdf}
\DeclareGraphicsExtensions{.eps}

\usepackage{amsmath}
\usepackage{amssymb}

\usepackage{cite}

\usepackage{nameref,hyperref}
\newcommand{\mynameref}[1]{#1}

\usepackage{microtype}
\DisableLigatures[f]{encoding = *, family = * }

\usepackage[labelfont=bf,labelsep=period]{caption}

\makeatletter
\renewcommand{\@biblabel}[1]{\quad#1.}
\makeatother

\graphicspath{{./figs/}}

\begin{document}

\title{Expectation-Maximization Binary Clustering\\ for Behavioural Annotation}
\author{
Joan Garriga$^{1}$, John R. Palmer$^{2}$, Aitana Oltra$^{1}$, Frederic Bartumeus$^{1,2,\ast}$ \\ \\
1 \textit{ICREA-Movement Ecology Lab (CEAB-CSIC),}\\ \textit{Cala Sant Francesc, 14, 17300, Blanes, Spain} \\
2 \textit{Centre for Ecological Research and Forestry Applications (CREAF)},\\ \textit{Cerdanyola del Vall\`es, 08193, Barcelona, Spain} \\
$\ast$ \textit{E-mail: fbartu@ceab.csic.es}
}

\maketitle

\begin{abstract}

  The growing capacity to process and store animal tracks has spurred the development of new methods to segment animal trajectories into elementary units of movement. Key challenges for movement trajectory segmentation are to (i) minimize the need of supervision, (ii) reduce computational costs, (iii) minimize the need of prior assumptions (\textit{e.g.} simple parametrizations), and (iv) capture biologically meaningful semantics, useful across a broad range of species. We introduce the Expectation-Maximization binary Clustering (EMbC), a general purpose, unsupervised approach to multivariate data clustering. The EMbC is a variant of the Expectation-Maximization Clustering (EMC), a clustering algorithm based on the maximum likelihood estimation of a Gaussian mixture model. This is an iterative algorithm with a closed form step solution and hence a reasonable computational cost. The method looks for a good compromise between statistical soundness and ease and generality of use (by minimizing prior assumptions
  and favouring the semantic interpretation of the final clustering). Here we focus on the suitability of the EMbC algorithm for behavioural annotation of movement data. We show and discuss the EMbC outputs in both simulated trajectories and empirical movement trajectories including different species and different tracking methodologies. We use synthetic trajectories to assess the performance of EMbC compared to classic EMC and Hidden Markov Models. Empirical trajectories allow us to explore the robustness of the EMbC to data loss and data inaccuracies, and assess the relationship between EMbC output and expert label assignments. Additionally, we suggest a smoothing procedure to account for temporal correlations among labels, and a proper visualization of the output for movement trajectories. Our algorithm is available as an R-package with a set of complementary functions to ease the analysis.

\end{abstract}

\newpage
\tableofcontents
\newpage

\section{Introduction}

Current movement research is undergoing a revolution. The growing capacity to collect high-resolution spatio-temporal movement data and radical improvements in data management and processing are unprecedented in this field and reminiscent of the bioinformatics revolution of genomics and proteomics two decades ago~\cite{Nathan2008}. A key challenge now is the analysis of massive movement datasets with largely different sampling rates and accuracies (e.g. high resolution GPS, standard GPS, Argos satellite, geolocation). In particular, it is critical to identify, in an unsupervised manner, movement trajectories' functional units, known as behavioural modes~\cite{Nathan2008, Giuggioli2010, Morales2010}. The behavioural mode is to the movement trajectory what gene is to the DNA sequence~\cite{Nathan2008a}.

Animal movement can be analysed as a set of measurable behavioural responses to a combination of internal states, environmental factors, and evolutionary/biological constraints. Such behavioural responses or modes are manifestations of the organism's decision mechanism, providing information about the cognitive process driving the movement~\cite{Thiebault2013}. Identifying behaviourally significant movement modes is crucial to bringing research beyond mere statistical descriptions of movement patterns and unravelling the underlying biological processes that determine animals movement and behaviour. Establishing robust connections between patterns and processes is important for the biological interpretation of the data but also for nurturing models of movement with larger predictive capacity.

Classic approaches to movement trajectory segmentation focus on the trajectory's structure by using local measures based on tortuosity \cite{Benhamou2004}, first-passage time \cite{Fauchald2003}, residence time \cite{Barraquand2008}, and positional entropy \cite{Roberts2004,Guilford2004}. Other relatively simple procedures include the use of cumulative sums methods \cite{Knell2012, Thiebault2013}. On the other side, more sophisticated procedures involve Bayesian estimates of the space-time probability of being in a given behavioural mode or state \cite{Jonsen2007, Bailey2008}. These can include location errors as well as environmental information \cite{Morales2004, Forester2007, Ovaskainen2008, Dalziel2008}. Recent examples clearly show the suitability of state-space models for estimation and inference of behavioural modes. Especially promising have been hidden Markov models (HMMs) and some HMM variants considering autocorrelations among variables or context-dependent transition probabilities \cite{
Bestley2010, Dean2013, Freeman2013, Joo2013, Charles2014, Gloaguen2015} as well as some multi-state models (MSM) \cite{Jackson2011, vanGils2015}. State-space approaches provide mechanistic and statistically sound insights about movement patterns but rely on strong \textit{a priori} assumptions about the underlying distributions governing movement states and state transitions in time, usually requiring species-specific and time-consuming parameter estimations. Indeed, there are several frameworks for state-space modelling and the criteria to identify the best framework for a given problem still remain unclear \cite{Tremblay2009}. Behavioural Change Point Analysis \cite{Gurarie2009} or t-Stochastic Neighbouring Embedding (tSNE) \cite{Maaten2009, Berman2014} do not require as many prior assumptions as state-space modelling but may also be limited by the fact that behaviours are described in a continuous parameter space which is not easy to interpret or discretize into behavioural units or modes. Many of the
current behavioural annotation procedures require intense computational resources or heavy data-specific supervision (\textit{e.g.} \cite{Dean2013, Freeman2013, Berman2014}) limiting its use with massive amounts of data or in comparative ecology studies (\textit{i.e.} patterns across different populations, species or tracking methodologies).

Among the future challenges for behavioural annotation of movement trajectories is to devise scalable and minimally supervised methods capable of keeping results understandable on the basis of a sufficiently generalized and robust statistical methodology. With this aim we here develop a generalized, computationally efficient method to identify behavioural modes in movement trajectories. The method is based on a minimally-supervised multi-variate clustering algorithm that takes into account both the correlations and the uncertainty of the variables (input features), making sense of multiple time-scale behavioural events. The underlying assumption is that behavioural modes can be described by a mixture of Gaussian distributions over a binary partition of the input space. Other assumptions are just aimed at minimizing biases and sensitivity to initial conditions. The method stands out in accomplishing a good compromise between the statistical significance and the biological interpretability (semantics) of the
output.

In the following we introduce the basic EMC framework and the EMbC variant for behavioural annotation, both its main conceptual features and implementation. Next we compare the EMbC with the EMC framework and basic HMM (using synthetic datasets) and with supervised expert labelling (using empirical datasets). The aim of these comparisons is not to rank the methods but simply to illustrate their relative strengths and weaknesses. Based on our results, we finally discuss the main features of the EMbC and we clarify when and why the EMbC might be most useful in the context of behavioural annotation in comparison to similar approaches.

\section{Models}

Gaussian mixture models are not new in animal movement modelling. As an example, the Gaussian mixture assumption is comparable to the current use of composite Brownian motion in mechanistic models of search \cite{Jansen2012, Reynolds2012}. Also, modelling approaches that assume Gaussian mixtures for movement variables such as speed have already proved useful for classifying animal tracking data into discrete movement modes \cite{Guilford2008, Freeman2010}. The novelty here is to use Gaussian mixtures into an EMC framework to behaviourally annotate movement trajectories in a meaningful way.

\subsection{Expectation-Maximization Clustering}

The Expectation-Maximization (\textit{EM}) algorithm \cite{Hartley1958, Dempster1977} is a well sounded, general, and iterative procedure for the maximum likelihood estimate of a parametric distribution underlying some given data, the latter eventually incomplete or showing missing values. A particular case of this algorithm is the parameter estimation of a Gaussian Mixture Model (GMM) when the generating Gaussian of each observation is unknown, commonly known as \textit{Expectation-Maximization Clustering} (EMC), a well known methodology for the identification of clusters (\textit{i.e.} different classes or patterns) in a multivariate data set.

The EMC formal statement is the following:

\begin{itemize}
 \item given a dataset $X=\left\lbrace\boldsymbol{x}_1,\,\dots,\,\boldsymbol{x}_n\right\rbrace$, where each data point $\boldsymbol{x}_i=\left( x_i^{\left(1\right)},\dots,x_i^{\left(m\right)}\right),\,\left(1\leq i\leq n\right)$, is a vector of values corresponding to $m$ variables, the EMC fits a $k$ multivariate-Gaussian Mixture Model defined by the parametric set $\Theta=\left\lbrace\boldsymbol\mu_{1},\boldsymbol\Sigma_{1},\pi_{1},\dots, \boldsymbol\mu_{k},\boldsymbol\Sigma_{k},\pi_{ k}\right\rbrace$, where $\boldsymbol\mu_{j},\boldsymbol\Sigma_{j},\pi_{j},\,\left(1\leq j\leq k\right)$, are respectively the vector of means, the covariance matrix and the mixing coefficient of multivariate Gaussian $j$. \end{itemize}

EMC is a two step iterative optimization method to estimate $\Theta^{*}$, alternating between estimates of the probability of a particular observation belonging to each cluster, and estimates about the parameters $\Theta$ that maximize the likelihood of these probabilities. For a GMM, the maximization equations have a forward analytical solution \cite{McLachlan1997, Gupta2010} that greatly simplifies the optimization procedure. In a few words, the algorithm proceeds as follows:

\begin{enumerate}
 \item Initialization: take a guess $\Theta^{g}$ over the set of parameters;
 \item Iteration loop: given the current guess $\Theta^{g}$,
 \begin{enumerate}
 \item Expectation step: for each data point $i$ and each cluster $j$, compute the likelihood weight $w_{ij}$, (a \textit{posterior} probability), of $\boldsymbol{x}_{i}$ being generated by Gaussian $j$, given by,

\begin{equation}\label{eq:lkhWeights}
 w_{ij}\equiv p\left(y_{i}=j\mid
\boldsymbol{x}_{i},\Theta^{g}\right)=\frac{\mathcal{N}\left(\boldsymbol{x}_{i}
\mid\boldsymbol\mu_{j},\boldsymbol\Sigma_{j}\right)\,\pi_{j}}{\sum_{j=1}^{k}\pi_
{j}\,\mathcal{N}\left(\boldsymbol{x}_{i}\mid\boldsymbol\mu_{j},
\boldsymbol\Sigma_{j}\right)}
\end{equation}

\noindent where $\mathcal{N}\left(\boldsymbol{x}_{i}\mid\boldsymbol\mu_{j},\boldsymbol\Sigma_{j}\right)$ is the multivariate Gaussian density function:

\begin{displaymath}
 \mathcal{N}\left(\boldsymbol{x}_{i}\mid\boldsymbol\mu_{j},\boldsymbol\Sigma_{j}
\right)=\frac{1}{\left(2\,\pi\right)^{m/2}\left|\boldsymbol\Sigma_{j}\right|^{
1/2}}\,e^{-\frac{1}{2}\left(\boldsymbol{x}_i-\boldsymbol\mu_{j}
\right)^T\boldsymbol\Sigma_j^{-1}\left(\boldsymbol{x}_i-\boldsymbol\mu_{j}
\right)}
\end{displaymath}

 \item Maximization step: compute a new set of parameters $\Theta^{new}$ that maximizes the likelihood~\cite{McLachlan1997} of these weights, given by the expressions,

\begin{eqnarray}
 \label{eq:maximization1}
 \pi_{j}^{new}&=&\frac{1}{n}\sum_{i=1}^{n}w_{ij}
 \\
 \label{eq:maximization2}
 \mu_j^{\left(l\right),new}&=&\frac{\sum_{i=1}^{n}w_{ij}\,x_i^{\left(l\right)}}{
\sum_{i=1}^{n}w_{ij}}
 \\
 \label{eq:maximization3}
 \sigma_j^{\left(r,s\right),new}&=&\frac{\sum_{i=1}^{n}w_{ij}\,\left(x_i^{
\left(r\right)}-\mu_{j}^{\left(r\right),new}\right)\,\left(x_{i}^{\left(s\right)
}-\mu_{j}^{\left(s\right),new}\right)}{\sum_{i=1}^{n}w_{ij}}
\end{eqnarray}

\noindent where $\mu_j^{\left(l\right),new},\, 1\leq l\leq m$, are the components of the mean vector $\boldsymbol\mu_j^{new}$, and $\sigma_j^{\left(r,s\right),new},\, 1\leq r\leq m,\, 1\leq s\leq m$ are the variances and covariances of the covariance matrix $\boldsymbol\Sigma_j^{new}$.

 \item Use $\Theta^{new}$ as the parametric guess for the next iteration, that is, take $\Theta^{g}\equiv\Theta^{new}$.

 \end{enumerate}

 \item Output classification: at the end of the process, each data point is assigned to its most probable cluster.

\end{enumerate}

This iterative procedure is theoretically guaranteed to increase the likelihood at each step and, although the algorithm does not promise to reach a global maximum of the likelihood function, it is indeed guaranteed to converge to a local maximum, dependent on the initial conditions \cite{McLachlan1997, McLachlan2004, Bilmes98}. In practice, it is common to start it from multiple random initial guesses and select the one with the largest likelihood. Usually, the process is stopped after a prefixed number of iterations or when the increments of likelihood are less than a prefixed $\delta$.

In a typical EMC implementation, the number of desired output clusters must be specified, and the algorithm will return that number of clusters. A value $\sigma_{min}^{\left(r,r\right)}>0$, $\left(1\leq r\leq m \right)$ must be specified to limit the minimum variance of each variable. This parameter avoids errors derived from indefinite covariance matrices along the optimization process. In practical terms, $\sigma_{min}$ can be directly related to measurement errors (or maximum resolution) of the variables and will limit the minimum range of variability (\textit {i.e.} minimum standard deviation) within the clusters obtained.

\subsection{EMbC Algorithm}

The \textit{generalized EM algorithm} is a family of variants of the EM algorithm aimed at overcoming particular problems (e.g. difficult E-step/M-step computations, slow convergence \cite{McLachlan2004}). The general behaviour of these variants is not always clear and they may not yield monotonic increases in the log likelihood over iterations \cite{McLachlan2004}. The \textit{Expectation-Maximization binary Clustering} (EMbC) algorithm is a variant of the EMC algorithm \cite{Hartley1958, Dempster1977} aimed to address: (i) clustering interpretability and, (ii) the variability in data reliability, two key issues in behavioural annotation of movement. The novelty is that the clustering is driven towards a statistically meaningful classification that should be easier to interpret by experts and that, similarly to other methods \cite{Kim2007, Tariquzzaman2012} it can take into account uncertainties associated to the data points.

\subsubsection{Clustering semantics: the delimiters}

In any unsupervised clustering procedure, one should distinguish cluster identification from cluster \textit{semantics}, the intuitive interpretation of the obtained clusters. Classical implementations of the EMC can generate statistically sound clustering configurations that are difficult to interpret in behavioural terms, that is, at the cost of clear semantics.

In the EMbC algorithm semantically meaningful clustering is achieved by introducing a set of parameters, denoted as \textit{delimiters}. A delimiter is a value that splits the range of a variable into a binary discretization. The whole set of delimiters defines a partition of the variable space into regions where each variable takes either low (L) or high (H) values. The binary nature of this partition is what favours the link between elementary and semantically meaningful labelling. As an example, classical behavioural annotation is commonly based on velocity and a turning behaviour estimate (\textit{e.g.} turning angle, angular correlation, tortuosity). In this case, a binary labelling has a direct intuitive interpretation: low velocities and low turns (LL) can be interpreted as \textit{resting}, low velocities and high turns (LH) as \textit{intensive search}, high velocities and low turns (HL) as \textit{travelling} or \textit{relocation}, and high velocities and high turns (HH) as \textit{extensive
search}. The semantic annotation is however variable-dependent and species-specific.

In a general multivariate case, each delimiter is associated to two adjacent clusters having the same combination of low and high values for all variables, except for the splitting variable, which takes low values in one and high values in the other. In other words, we have one delimiter for each variable and each combination of highs and lows of the rest of the variables. For $m$ variables this makes a total of $m\,2^{m-1}$ delimiters. We use a multivariate notation denoting delimiters by $r_{Z}$ where $Z$ is a length $m$ sub-index, based on an ordered sequence of the variables. Each element of the sub-index is either $L$ or $H$ except for the splitting variable for which we use a dot, according to the combination of values at both adjacent clusters. As an example, in a 3-variate case, $r_{L.H}$ denotes the delimiter for the second variable, separating the two clusters $LLH$ and $LHH$, in which the first and the third variables take low and high values respectively.

\begin{figure}[!p]
\includegraphics[width=15cm,height=6.5cm]{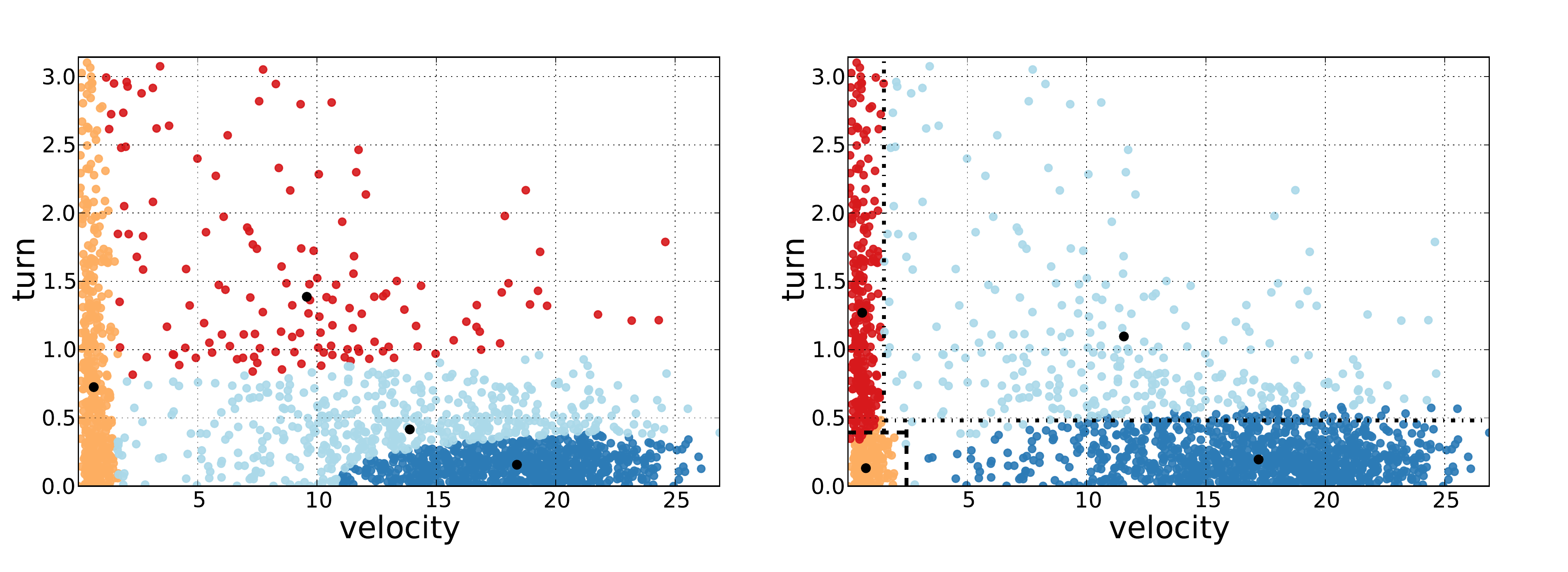}
\caption{\textbf{Cluster semantics}. Comparison of the EMC (left) and EMbC (right) algorithms. Bivariate (velocity/turn) scatter-plots showing the clustering reached by each algorithm, corresponding to the same trajectory and exact initial conditions. Clusters are shown in different colours. In the right panel, the EMbC delimiters determining the final L/H binary regions are depicted as dashed lines ($r_{.L},r_{L.}$) and dot-dashed lines ($r_{.H},r_{H.}$). The centroids of each cluster are shown as black dots. Left: the EMC yields an output clustering that is difficult to link to a clear semantics. Right: the EMbC is driven by the delimiters, forcing the centroids to lay within the associated binary regions, yielding a final clustering that can be clearly interpreted in terms of L/H values of the variables (orange:LL, red:LH, cyan:HL and blue:HH). The matching among binary regions and clusters is not perfect because data-points are assigned to clusters depending on their weights, not on the delimiter values. In this case, the EMbC performs better (the clustering log likelihoods are -3.3368 for the EMC and -3.2180 for the EMbC), but this result can not be generalized.}
\label{fig_01}
\end{figure}

Conceptually, the delimiters are related to the frontier of equiprobability between two adjacent clusters, and are used to bound the computation of the Gaussian means within the regions that they delimit. In such a way, the mean of each cluster can not drift away from its associated binary region. To illustrate this issue, we show a comparison of two bivariate (velocity and turning angle) clusterings of the same trajectory (Fig.~\ref{fig_01}): one resulting from a classical EMC implementation (left panel), and the other resulting from the EMbC variant, the dashed lines depicting the final value of the delimiters computed by the EMbC algorithm (right panel). Starting with exact initial conditions, these two algorithms yield output clusterings corresponding to different local optima. In the left panel it is difficult to obtain a clear semantics based on velocity/turn. In the right panel the clustering shows a meaningful partition of the variable space into LL/LH/HL/HH regions, accounting for a clear cluster
semantics.

\begin{center}
\textbf{Figure~\ref{fig_01}. Cluster semantics.}
\end{center}

Regardless of the values of the delimiters, the EMbC assigns data-points to the most probable cluster. This is the reason for the few mismatches that can be observed between the clusters and the binary regions in the right panel. Only in case of equal probabilities, the delimiters constitute a further criterion to assign labels to data-points. Importantly, there is no guarantee that either of the algorithms (EMC and EMbC) will always be better in terms of likelihood. Both EMC and EMbC will just reach the best optimum attainable from any given starting point. In our example, the local optimum based on EMbC is better (Fig.~\ref{fig_01}~right panel), but this must not be generalized. The concern here is not to reach higher likelihood partitions but rather to reach meaningful partitions even at some cost in likelihood.

\subsubsection{Data accuracy and reliability}

Movement trajectory data is associated to different sources of uncertainty: (i) spatial errors due to technical limitations of the systems used (\textit{e.g.} GPS, Argos) or interferences in signal transmissions of the geopositioning system; and (ii) temporal errors due to difficulties in inferring a location at a given time, which generates irregular time steps. Therefore, estimated variables such as velocity or turn, which depend on the sampling rate and on the locations themselves (Section~1 in \mynameref{S1 Text}), present different degrees of reliability or accuracy.

Similarly to \cite{Kim2007,Tariquzzaman2012}, the reliability of the data is implemented as an additional weighting coefficient in Equations \ref{eq:maximization2} and \ref{eq:maximization3}, giving less weight to the less accurate values in the estimation of the Gaussian parameters, and favouring the more accurate ones. These coefficients should be given by a \textit{reliability} function that can not be generalized, as it will be variable-specific and dependent on the source of error considered. For the general case we denote them as,

\begin{equation}\label{eq:certainty}
 u_i^{\left(l\right)} = \mathcal{E}(x^{\left(l\right)}_i)
\end{equation}

\noindent where $\mathcal{E}(x^{\left(l\right)}_i)$ is a function that returns a normalized reliability coefficient for the data value $x^{\left(l\right)}_{i}$ based on the source (or multiple sources) of error that might be operating upon the input variable $l$. In \mynameref{S1 Text} Section~3 we suggest an example of a reliability function that can be used to take into account the reliability of the values of velocity and turn computed from a real trajectory with heterogeneous time intervals.

\subsubsection{Implementation}\label{Implementation}

\begin{figure}[!p]
\includegraphics[width=15cm,height=12cm]{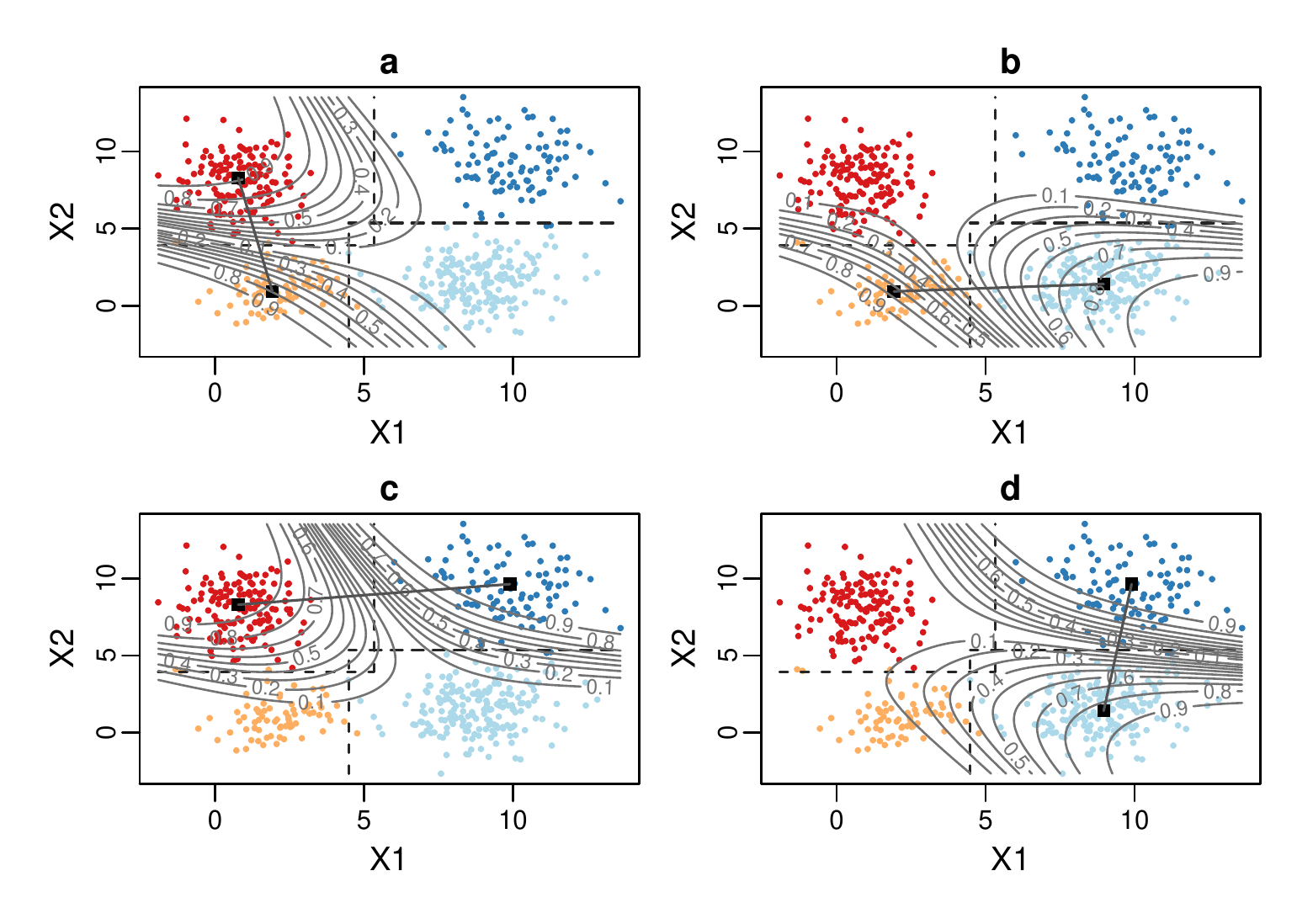}
\caption{\textbf{Computation of the delimiters}. This is a synthetic example with data drawn from a bivariate $\left(X1,\,X2\right)$ GMM with four components, showing the state of the clustering at the third iteration. Current delimiters are shown as dashed lines. To depict the idea, we defined a grid $\mathcal{G}$ over the range of the scatter-plot and we computed the likelihood weights $\boldsymbol{w}_{LL}\left(\mathcal{G}\right)$, $\boldsymbol{w}_{LH}\left(\mathcal{G}\right)$, $\boldsymbol{w}_{HL}\left(\mathcal{G}\right)$, $\boldsymbol{w}_{HH}\left(\mathcal{G}\right)$. We show the contour lines corresponding to the differences in likelihood weight and the lines connecting the means of the two adjacent clusters: panel a) shows the line $\left(\mu_{LL},\mu_{LH}\right)$ and the contour $abs\left(\boldsymbol{w}_{LL}\left(\mathcal{G}\right)-\boldsymbol{w}_{LH} \left(\mathcal{G}\right)\right)$; panel b) shows the line $\left(\mu_{LL},\mu_{HL}\right)$ and the contour $abs\left(\boldsymbol{w}_{LL}\left(\mathcal{G}\right)-\boldsymbol{w}_{HL} \left(\mathcal{G}\right)\right)$; panel c) shows the line $\left(\mu_{LH},\mu_{HH}\right)$ and the contour $abs\left(\boldsymbol{w}_{LH}\left(\mathcal{G}\right)-\boldsymbol{w}_{HH} \left(\mathcal{G}\right)\right)$; and panel d) shows the line $\left(\mu_{HL},\mu_{HH}\right)$ and the contour $abs\left(\boldsymbol{w}_{HL}\left(\mathcal{G}\right)-\boldsymbol{w}_{HH} \left(\mathcal{G}\right)\right)$. In each case, we can observe that the delimiter crosses the corresponding line between means at the point with minimum likelihood difference; panel a) $r_{L.}$ (turn splitting value for low values of speed); panel b) $r_{.L}$ (speed splitting value for low values of turn); panel c) $r_{.H}$ (speed splitting value for high values of turn); panel d) $r_{H.}$ (turn splitting value for high values of speed).}
\label{fig_02}
\end{figure}

\begin{figure}[!p]
\includegraphics[width=15cm,height=6.5cm]{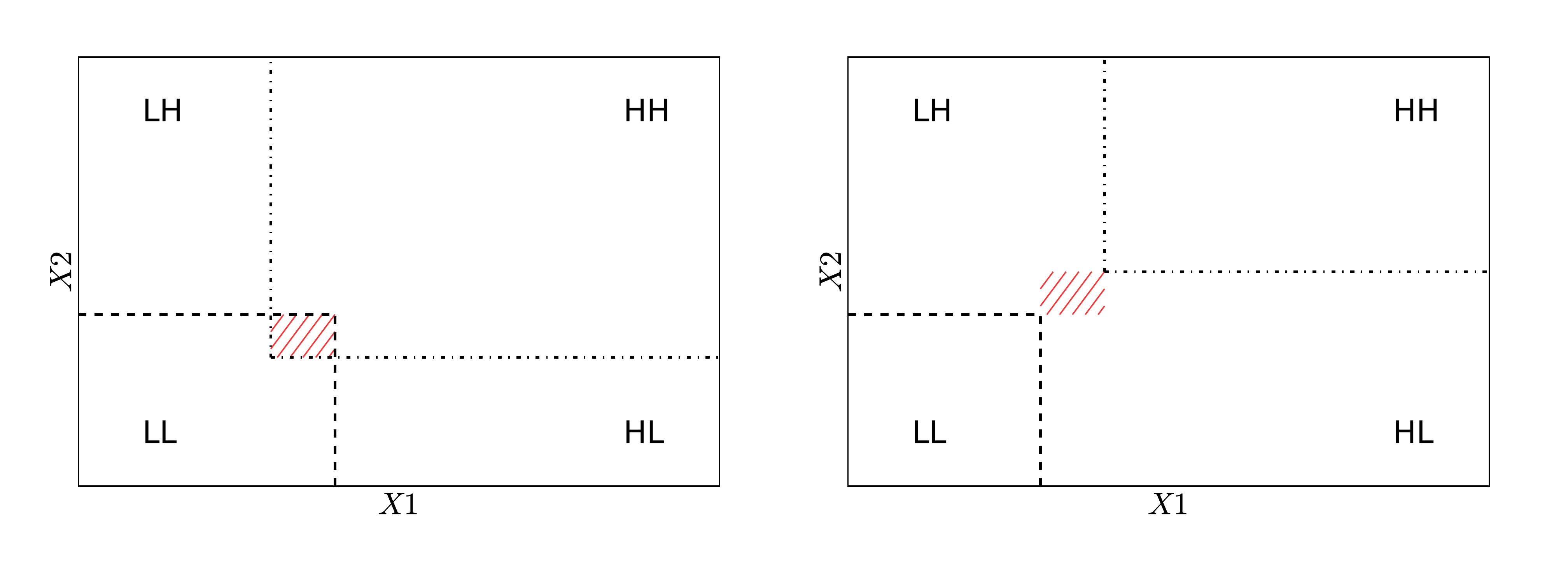}
\caption{\textbf{Definition of the binary regions.} At each iteration step, the most common situation is that the delimiters do not determine a perfect partition of the variable space. We show two typical cases for the bivariate $\left(X1,\,X2\right)$ case with delimiters overlapping (left panel) and non-overlapping (right panel). Delimiters are shown as dashed lines ($r_{.L},r_{L.}$) and as dot-dashed lines ($r_{.H},r_{H.}$). Left: the data points in the middle red area may belong either to $\mathcal{R}_{LL}$ or $\mathcal{R}_{HH}$, hence they are considered in the computation of both $\mu_{LL}$ and $\mu_{HH}$. Right: with non-overlapping delimiters, we can still figure out an overlapping area between regions $\mathcal{R}_{LH}$ and $\mathcal{R}_{HL}$ by extending the delimiters (middle red area), hence data points in this area are considered in the computation of both $\mu_{LH}$ and $\mu_{HL}$.}
\label{fig_03}
\end{figure}

Implementing the modifications described above to account for clustering semantics and data reliability imply relevant changes in the maximization step (M-step) of the algorithm:

\begin{enumerate}
 \item Foremost, the delimiters have to be computed. The values of the delimiters correspond to the point of minimum difference in likelihood weight in between two adjacent clusters. At each iteration the delimiters are computed by projecting the data points onto the straight line connecting the current means of the adjacent clusters. The likelihood weights of the projected data points are computed for both clusters with the current parameters, and the delimiter is set to the value of the data point for which the difference in those likelihoods is minimum (Fig.~\ref{fig_02}).

\begin{center}
\textbf{Figure~\ref{fig_02}. Computation of the delimiters.}
\end{center}

 \item For each cluster $j$, we must determine the set $\mathcal{R}_{j}$ of points lying in the region determined by the corresponding delimiters. We note that in the general case, the delimiters will not constitute a perfectly definite partition of the variables space, and some points may belong to different regions at the same time, as shown in Fig.~\ref{fig_03}, contributing to the computation of the Gaussian means of all related clusters.

\begin{center}
\textbf{Figure~\ref{fig_03}. Definition of the binary regions.}
\end{center}

 \item Finally, we recompute the Gaussian parameters, bounding equation \ref{eq:maximization2} to the sets $\mathcal{R}_{j}$ and including the reliability function in equations \ref{eq:maximization2} and \ref{eq:maximization3}, that is,

\begin{eqnarray}
 \label{eq:boundedMaximization2}
 \mu_j^{\left(l\right),new} &=&
\frac{\sum_{i\in\mathcal{R}_{j}}\,u_i^{\left(l\right)}\,w_{ij}\,x_i^{
\left(l\right)}}{
 \sum_{i\in\mathcal{R}_{j}}u_i^{\left(l\right)}\,w_{ij}}
 \\
 \label{eq:boundedMaximization3}
 \sigma_j^{\left(r,s\right),new} &=&
\frac{\sum_{i=1}^{n}u_i^{\left(r,s\right)}\,w_{ij}\,\left(x_i^{\left(r\right)}
-\mu_j^{\left(r\right),new}\right)\,\left(x_i^{\left(s\right)}-\mu_j^{
\left(s\right),new}\right)}{\sum_{i=1}^{n}u_i^{\left(r,s\right)}\,w_{ij}}
\end{eqnarray}
\end{enumerate}

\noindent where $u_i^{\left(r,s\right)}$ weights the combined effect of uncertainty on variables $r$ and $s$, and is computed as the normalized length,

\begin{displaymath}
u_i^{\left(r,s\right)}=\frac{1}{\sqrt{2}}\,\sqrt{\left(u_i^{\left(r\right)}
\right)^2+\left(u_i^{\left(s\right)}\right)^2}
\end{displaymath}

The delimiters become the essential part of the parametric set $\Theta$, and therefore, the model is no longer a standard GMM but a constrained variant. The optimization through the likelihood space is driven by the new conditions imposed, which force each Gaussian to have its mean inside meaningful regions, restricting the potential positions of the cluster centroids and the type of clusterings allowed.

A major consequence is that Equations \ref{eq:boundedMaximization2} and \ref{eq:boundedMaximization3} do not correspond to maximization expressions. This change however, does not jeopardize the convergence of the EMbC algorithm. The effect of our modifications in the EMC algorithm is an increasing likelihood optimization process, interspersed with likelihood drops at sporadic iterations. Every drop in likelihood can be considered a restart in the likelihood landscape from a new (but not so blind) guess, with the likelihood being lower compared to the previous step but higher compared to the likelihood value from which we started. A steady likelihood decrease at some stage of the optimization process is an indication of some discrepancy between the binary and the optimal likelihood partitions, and that the input data might not be suited for a binary partition. In such cases, the algorithm may get stuck in a cycle balancing between both possible solutions. This situation (more likely to occur on the last
iterations) is automatically detected and the algorithm stops returning a corresponding warning message. The likelihood dynamics are further discussed and illustrated with some examples in \mynameref{S1 Text} Section~5.

Unlike the EMC algorithm, the number of output clusters is given here by the number of variables used, $k=2^m$. However, during the process of likelihood optimization, some clusters can vanish while being absorbed by adjacent clusters. Thus, $k=2^m$ is only an upper bound to the final number of clusters. This limitation in the number of clusters is not a drawback but rather a consistency with our main motivation of favouring the semantic interpretation of the final clustering. Although there is no restriction on the number of variables (we are not considering computational limitations) the EMbC is intended to be used with not more than 5 or 6 variables, yielding a maximum of 32 or 64 clusters, what is usually far beyond the number of potential behaviours of interest in any biological application, and far beyond the number of behaviours that an expert might easily handle. The key point here is to determine a few variables conveying the right information to decode the set of behaviours of interest. It is worth
noting that we are not talking about automated feature selection or dimensionality reduction methodologies (\textit{e.g., principal component analysis}) as this would go against the interpretability of the output clustering which is the ultimate purpose of the EMbC algorithm. Input features should be selected based on their physical or biological meaning.

The parameter $\sigma_{min}$ is variable-specific and determines the minimal resolution of the clusters. It can be set by default to orders of magnitude lower than the expected variances (\textit{e.g} $\sigma_{min}=2.22e{-16}$ or whatever it is the double-precision of the computer) for each of the variables or else be used to limit the minimum range of variability (\textit{i.e.} minimum standard deviation) within the clusters. Rather than a subjective question, this is usually related to the physical concept expressed or measured by the variable under consideration. For instance, regarding to movement variables like velocity and turn, $\sigma_{min}$ can be directly related to physical/biological constraints as well as to measurement errors (or maximum resolution) of geolocation devices. Thus, values of $\sigma_{min}$ in the order of $0.01$ m/s for velocity and $0.087$ rad (5 degrees) for turns, would work for a wide range of species. For this reason, $\sigma_{min}$ should be regarded as a variable-specific
factor that fixes the analysis resolution rather than a user free parameter.

Also key in the algorithm implementation is to minimize prior assumptions, biases and sensitivity to initial conditions. With this aim, the EMbC starting point is automatically set as the most uninformative condition, that is: (i) each data point is assigned a uniform probability of belonging to each cluster, (ii) the prior marginal distribution of the clusters is also uniform (each cluster starts with the same number of data points), (iii) the starting partition, \textit{i.e.} the initial delimiters position, is selected based on a global maximum entropy criterion, thus conveying the minimum information possible. The latter condition is computed by sequentially selecting the variable such that its median value splits the related set of data into high and low subsets with maximum entropy. This is a simple algorithm for the 2D case but slightly more complex for the general case of $d$ dimensions.

\section{Analysis}

We use simulated and empirical trajectories to asses the EMbC algorithm and illustrate its main features. Our examples are mostly based on local measures of velocity and turn but we also show an example with other movement variables (Sections~1~and~2 in \mynameref{S1 Text}).

Synthetically generated and annotated movement trajectories allow us to measure the performance of the algorithm and compare it with closely related methods such as EMC \cite{Hartley1958, Dempster1977} and HMM, commonly used for modelling of animal movement data \cite{Dean2013, Langrock2012, Joo2013}. We use the implementations of EMC and HMM included in the R-packages EMCluster~\cite{Chen2012EMClusterpackage, Chen2012EMClustervignette} and DepmixS4~\cite{depmixS4}, respectively.

Synthetic trajectories are generated by assuming four clusters (behavioural modes) with different degrees of mixture or overlap, $\gamma=\left\lbrace0.01,0.05,0.1\right\rbrace$, where the lower the value of $\gamma$ the more blurred are the clusters. The trajectories are of different lengths $n=\left\lbrace50,100,200,400,800,1600\right\rbrace$ and the sequence of behavioural modes or states is constructed either by sampling states from a $4\times4$ transition matrix (Markov-chain sampling) or else by sampling states using the parameters of the prior distributions $\pi_{j}$\,,$1\leq j\leq 4$ (prior-mixture sampling), (see Section~6 in \mynameref{S1 Text} for more details).

Our empirical tracks (see Table~\ref{tbl:empiricalData}) cover different ecological contexts and a variety of tracking technologies including high-resolution GPS (shearwater), standard GPS (bat), Argos (osprey) and video recorded (nematode) datasets. The corresponding data sets are further described in \mynameref{S2 Text} and are included in \mynameref{S3 Data}.

\begin{table}[!ht]
\caption{\textbf{Empirical datasets.}}
\end{table}

EMbC outputs are shown in different ways (\textit{i.e.} scatter-plots, labelling profiles) including a bursted visualization of annotated trajectories based on the conversion into segments of all consecutive locations sharing the same label (\mynameref{S1 Text} Section~4). In the case of supervised datasets (\textit{i.e.} synthetic datasets or empirical datasets with expert labelling), we use confusion matrices to yield a numerical assessment of the performance of the algorithm with respect to the supervised labelling. Commonly used performance measures based on the confusion matrix are \textit{recall}, \textit{precision} and \textit{F-measure} (\mynameref{S1 Text} Section~7). However, in the case of empirical datasets, confusion matrices can not be considered a strict numerical assessment of performance because expert labelling might be based on information (\textit{e.g.} visual information) not conveyed by the input features used with the EMbC (\textit{i.e.} velocity and turn in our case).However, by
quantifying how EMbC annotation is redistributed into expert label assignments one can gain much knowledge on the functioning of the EMbC algorithm itself.

Other aspects analysed are: (i) the coarse-graining of EMbC behavioural annotation, and (ii) the robustness of the EMbC with respect to potential sources of error like data loss and data inaccuracy.

The results that we show are mostly direct outputs of the EMbC R-package. In the S2 text file we spell out the code used to generate them and we work through them further to give a brief overview of the use of the package. The empirical data sets used in the examples are included in \mynameref{S3 Data}.

\subsection{Smoothing of annotated trajectories}

The EMbC algorithm generates \textit{local} behavioural annotations without considering their temporal context. Based on EMbC, labels are given for each observed location and reveal any small change in behaviour irrespective of how this change is framed in a broader temporal context (\textit{e.g.} a long-term predominant behavioural mode). If coarse-grained patterns are desired, the EMbC provides two means for smoothing the output:

\begin{enumerate}
 \item Pre-processing of the trajectory using running windows to compute averaged local measures, with the length of the running window representing a behavioural scale of interest.
 \item Post-processing of the output based on the temporal behavioural correlations, a feature explicitly implemented in state-space segmentation algorithms \cite{Jonsen2007, Bailey2008, Dean2013, Freeman2013}.
\end{enumerate}

In the latter case, the EMbC smoothing function makes use of the likelihood weights $w_{ij}$ of location $i$ belonging to cluster $j$, information provided by the EMbC algorithm. In its most basic implementation, the function looks for \textit{singles}, that is, locations with labels that differ from equally labelled neighbouring locations, and checks the condition $\left(w_{ic}-w_{in}\right)\leq\delta_{w}$, where $i$ is the \textit{single} location index, $w_{ic}$ and $w_{in}$ are the likelihood weights with respect to its current and its neighbouring assignments (clusters $c$ and $n$, respectively), and $\delta_{w}$ is a threshold parameter expressing the user's will to accept the change. The subjectiveness of this parameter is our reason for keeping this smoothing function apart from the overall clustering procedure.

\subsection{Robustness to data loss and data inaccuracy}

\begin{figure}[!p]
\includegraphics[width=15cm,height=6.5cm]{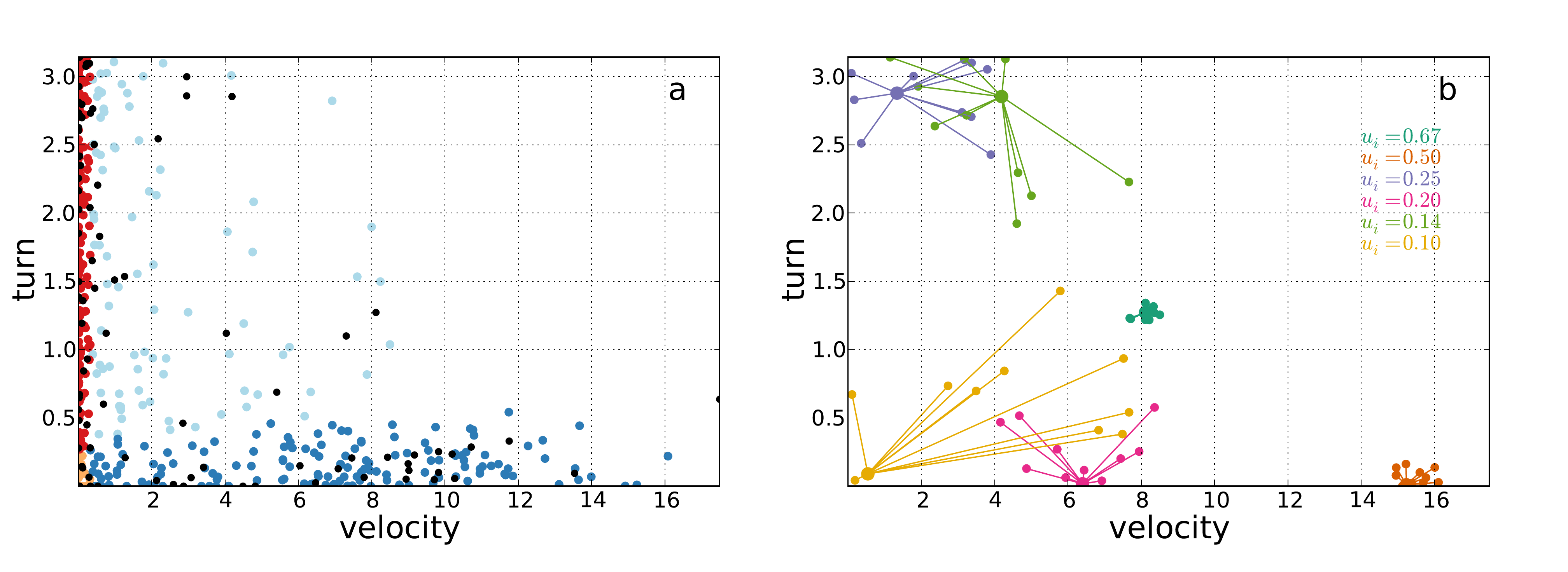}
\caption{\textbf{Procedures for robustness tests.} Procedures used in the robustness tests. a) Data-loss: sub-sampled datasets are generated by assigning a uniform random value $0\leq p_{i}\leq 1$ to each data-point and removing all those points with $p_i<k_{dl}$, with $0\leq k_{dl}\leq 1$ being the data loss factor, (in this example $k_{dl}=0.2$, removed points are marked as black dots). b) Data-inaccuracy: jittered datasets are generated by jittering the data-points using a noise function based on the associated uncertainties (Equations~\ref{eq:jitteringValue} and~\ref{eq:jitteringNoise}); we show some example data points connected with several jittered versions of themselves with $k_{di}=0.05$, using different colours to identify the correspondences, and also to relate each one with its associated reliability $u_{i}$ indicated in the legend; note that the more unreliable is a data point the more different could have been its observed value.}
\label{fig_04}
\end{figure}

We studied the robustness of EMbC annotation to data loss by removing data points from the set of velocity/turn pairs (Fig.~\ref{fig_04}a). Note that we cannot study data loss by eliminating locations from the trajectory because this would simply change the values of velocity and turn (which depend on actual sampling intervals), leading to a totally \textit{new} dataset. In the sub-sampling process it is important to preserve the underlying behavioural distribution. Thus, sub-sampled datasets were generated by assigning a uniform random value $0\leq p_i\leq1$ to each data point and removing all those points with $p_i<k_{dl}$, with $0< k_{dl}\leq 1$ being $k_{dl}$ the data loss factor. For each empirical trajectory, we generated datasets with different values of $k_{dl}$. After running the EMbC on the subsampled datasets we compared the output labels with the corresponding labels in the original (full) dataset, the latter considered the ground truth for comparative purposes.

We also explored the effect of including the reliability of the data (Equation~\ref{eq:certainty}) in Equations \ref{eq:boundedMaximization2} and \ref{eq:boundedMaximization3}. In particular, we considered the effect of sampling rate heterogeneity (\textit{i.e.} the larger the time gap between two successive locations, the larger the probability of inaccurate velocity and turn estimates), and to what extent our approach decreased the sensitivity of the final clustering to this source of inaccuracy (see Equation~10 in \mynameref{S1 Text} Section~3 as an instance of Equation~\ref{eq:certainty} devised for this particular case). We did so by jittering the data points in the scatter plot (Fig.~\ref{fig_04}b) using a noise function based on a uniform distribution over an area around the data point proportional to the associated time gap,

\begin{equation}\label{eq:jitteringValue}
 max\left(min\left(X\right),\boldsymbol{x}_{i}-\boldsymbol{\Delta}_{i}\right)<\boldsymbol{\hat{x}}_i<min\left(max\left(X\right),\boldsymbol{x}_{i}+\boldsymbol{\Delta}_{i}\right)\quad,
\end{equation}

\noindent where $X$ is the (multivariate) dataset, $\boldsymbol{x}_{i}$ and $\boldsymbol{\hat{x}}_{i}$ (vectors) are the original and jittered data points, and $\boldsymbol{\Delta}_{i}$ (vector) is computed as,

\begin{equation}\label{eq:jitteringNoise}
 \boldsymbol{\Delta}_{i}=
 k_{di}\,max\left(X\right)\;
 \left(\tau_{i}-\tilde{\tau}\right)/\tilde{\tau}\quad,
\end{equation}

\noindent where $0< k_{di} \ll 1$ is a data inaccuracy factor determining a jittering range $k_{di}\,max\left(X\right)$. Thus, $\boldsymbol{\Delta}_{i}$ is a fraction of the jittering range proportional to the relative length of the time interval $\tau_{i}=t_{i+1}-t_{i}$ with respect to the most frequent time interval $\tilde{\tau}$ (the mode of the $\tau$ distribution). For each empirical trajectory, and for different $k_{di}$ values, we compared the labellings obtained in jittered datasets with the corresponding non-jittered labellings, with and without implementing a reliability function in Equations \ref{eq:boundedMaximization2} and \ref{eq:boundedMaximization3}. This is only a particular example focused on inaccuracies derived from sampling heterogeneity but the same could apply to other sources of uncertainty, such as geopositioning errors. The effects would be similar, since the higher the uncertainty of the values, the less their influence in determining the final clustering.

\begin{center}
\textbf{Figure~\ref{fig_04}. Procedures for robustness tests.}
\end{center}

\section{Results}

We used synthetic trajectories and empirical datasets to evaluate the performance and illustrate the outputs of the EMbC algorithm. The sequences of movement states generated in the synthetic trajectories come from two sampling schemes: Markov-chain and mixture-prior. The empirical datasets covered different tracking technologies (e.g. GPS, Argos) and a wide range of sampling heterogeneity.

\subsection{Simulated trajectories}

\begin{figure}[!p]
\includegraphics[width=15cm,height=12cm]{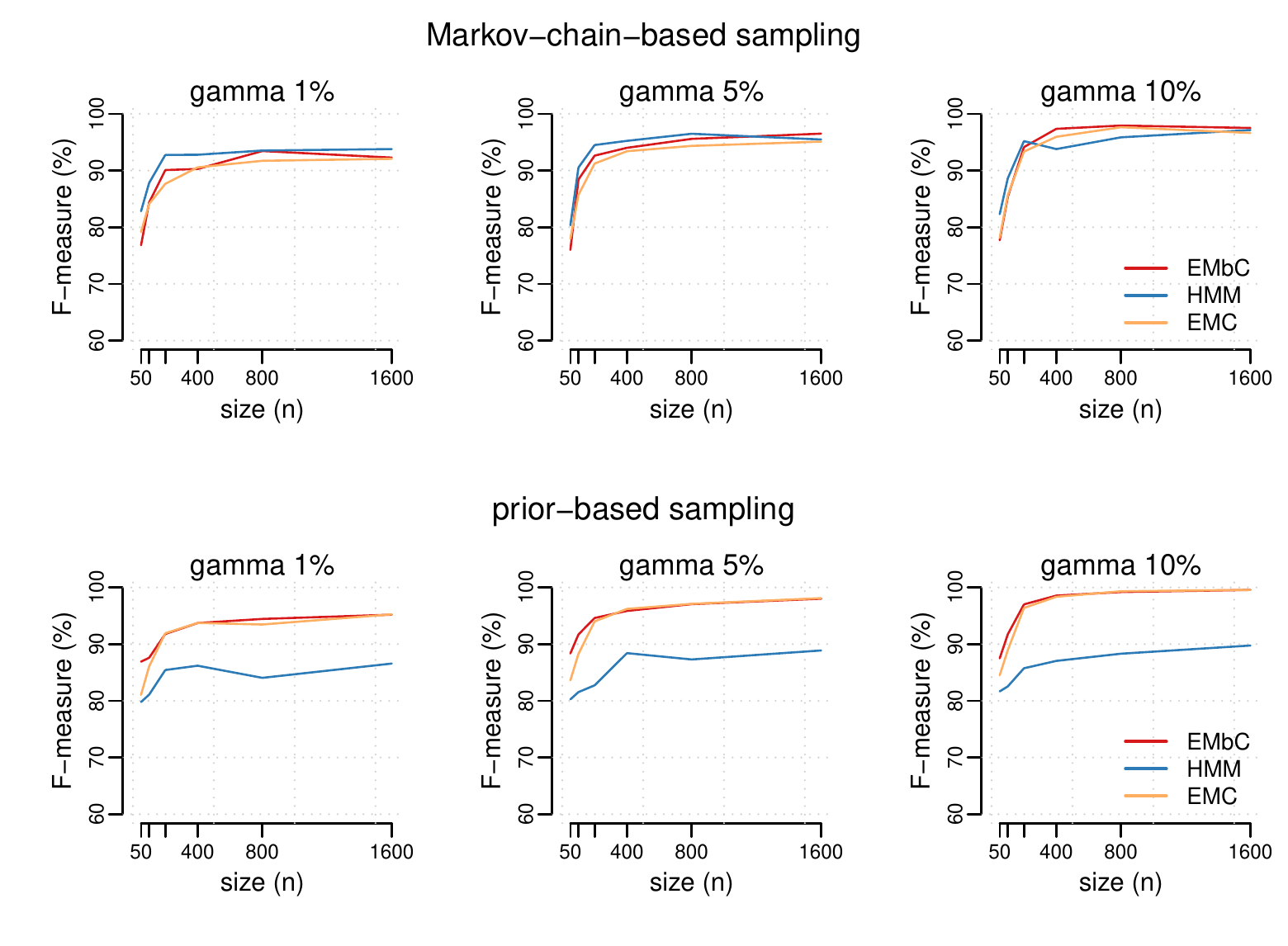}
\caption{\textbf{Performance comparison among EMbC, EMC, and HMM.} Average performance of EMbC, EMC and HMM on 100 synthetic trajectories drawn from a GMM (4 components) using a Markov-chain-based sampling scheme (top panel) and a prior-based sampling scheme (bottom panel), for different trajectory sizes ($n=\left\lbrace 50,100,200,400,800,1600\right\rbrace$) and definition of the binary regions ($\gamma=\left\lbrace 0.01,0.05,0.10\right\rbrace$). Values of performance are given in terms of F-measure.}
\label{fig_05}
\end{figure}

The EMbC algorithm recovered the modelled clusters but with some expected sensitivity to both the level of mixture of the clusters $\gamma$, and the size of the data set $n$. In general (Fig.~\ref{fig_05}), the performance was above 90\% for $n\geq 200$ decreasing around 80\% for the shortest trajectories, \textit{i.e.} $n\leq 100$.

\begin{center}
\textbf{Figure~\ref{fig_05}. Performance comparison among EMbC, EMC, and HMM.}
\end{center}

With synthetic trajectories derived from the Markov-chain sampling scheme, where the sequence of states comes from a transition probability matrix, (Fig.~\ref{fig_05} upper panels), the three algorithms (EMC, EMbC and HMM) showed a similar behaviour. For $\gamma=\left\lbrace0.01,0.05\right\rbrace$ (well-mixed clusters) the performance of the EMbC was in between the one of the HMM (the best) and the EMC. However, for $\gamma=0.1$ (well-defined clusters) and $n\geq 200$ the EMbC outperformed the HMM. Compared to EMC and EMbC, HMM works best when the binary partitions are blurred but the temporal sequence of states is well-defined, according to a transition matrix, and it can adequately exploit this information to improve state assignment.

With synthetic trajectories derived from the mixture-prior sampling scheme, where the sequence of states comes from the set of prior cluster distributions (Fig.~\ref{fig_05} lower panels), the EMbC and the EMC presented similar results. Expectation-Maximization clustering procedures do not take into account the temporal correlation of states but take the best out of the synthetically generated binary partitions, even if the clusters are well-mixed or blurred. The EMbC performed slightly better than EMC for low values of $\gamma$ and $n$. In contrast, the performance of the HMM was clearly much lower, both compared to EMC and EMbC, and also compared to the results obtained when HMMs are applied to trajectories based on Markov-chain sampling schemes.

The larger the size of the data set, the more evidence about the partition of the input space, and the better the performance of the three algorithms for both sets of trajectories (Markov-chain and mixture-prior sampling schemes). Tables~2 and~3 in \mynameref{S1 Text} reinforce the idea that EMbC performs better when the information is compromised, either because the clusters are not well-defined (small $\gamma$s) or because the amount of information is small (low $n$s). In addition, the EMbC leads straightforwardly to the binary partition and keeps a high stability in the results with the lowest values of \textit{root mean square error} (RMSE), (see Tables~2 and~3 in \mynameref{S1 Text}). In contrast, the EMC and the HMM algorithms are extremely sensitive to the starting conditions and they do not guarantee an output binary partition despite of the underlying binary clustered distribution of the input data (\textit{i.e.} one has to carry out multiple runs using different starting seeds and check whether the
final partition is a binary one or not, otherwise it makes no sense to compare the output with the state labels).

\begin{figure}[!p]
\includegraphics[width=15cm,height=17cm]{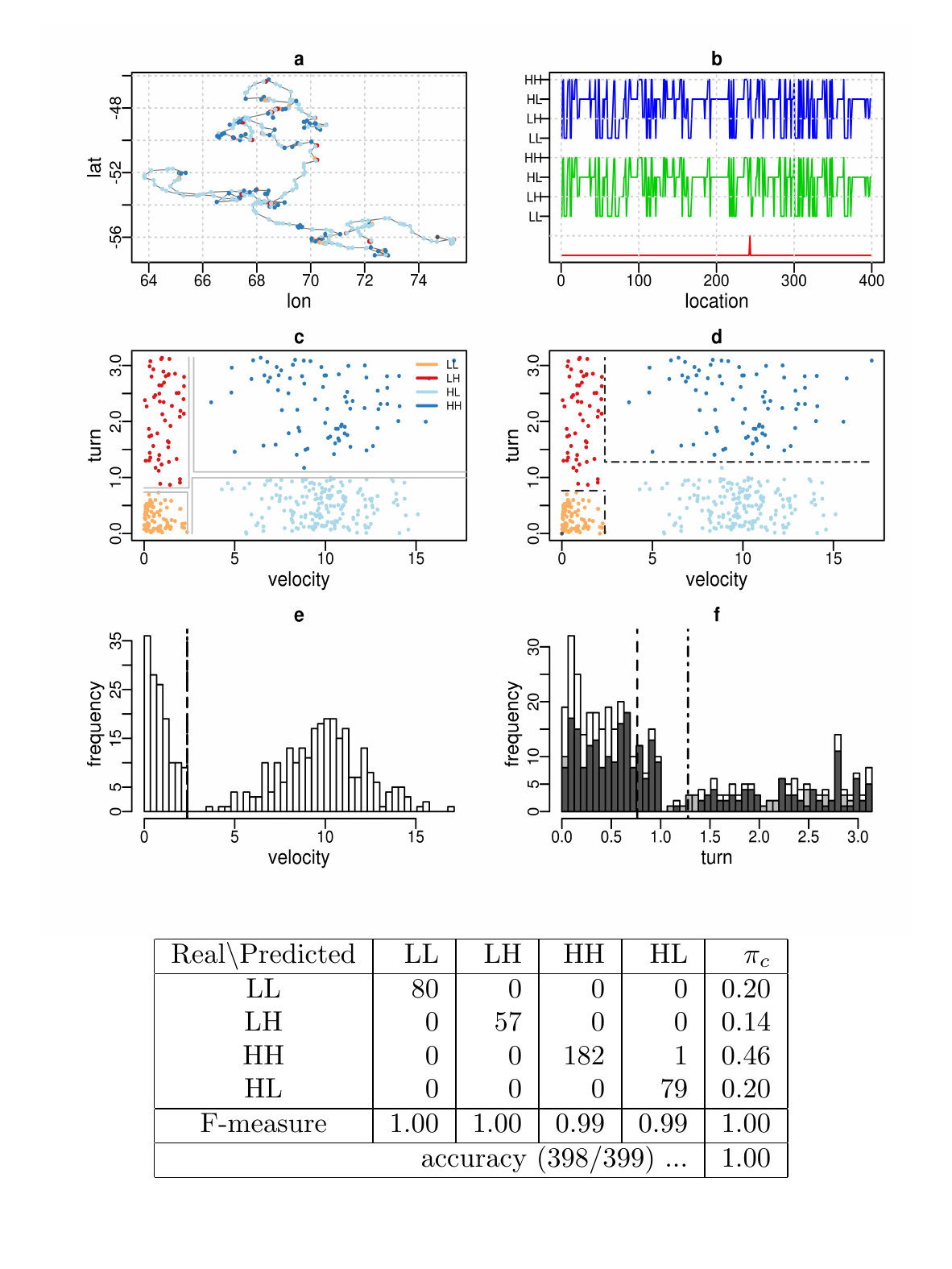}
\caption{\textbf{Synthetic trajectory.} Simulated trajectory with $N=400$ and $\gamma=0.05$. Panels: (a) trajectory grid plot, (b) temporal behaviour profile with output labelling (top), reference labelling (centre) and labelling differences between them (bottom), (c) reference velocity/turn scatter plot showing the limits of the binary regions (grey lines), (d) output velocity/turn scatter-plot showing the resulting delimiters $r_{.L},r_{L.}$ (dashed lines) and $r_{.H},r_{H.}$ (dot-dashed lines), (e) velocity, and (f) turning angle frequency distributions (white colour). The turn distribution for low/high values of velocity is shown separately in light/dark grey colours, respectively. Bottom: Confusion matrix and performance measures.} \label{fig_06}
\end{figure}

As an example, Fig.~\ref{fig_06} shows the EMbC output for a Markov-chain sampled trajectory ($n=400$, $\gamma=0.05$), where the clusters were perfectly recovered.

\begin{center}
\textbf{Figure~\ref{fig_06}. Synthetic trajectory.}
\end{center}

\subsection{Empirical trajectories}

\begin{figure}[!p]
\includegraphics[width=14cm,height=15.5cm]{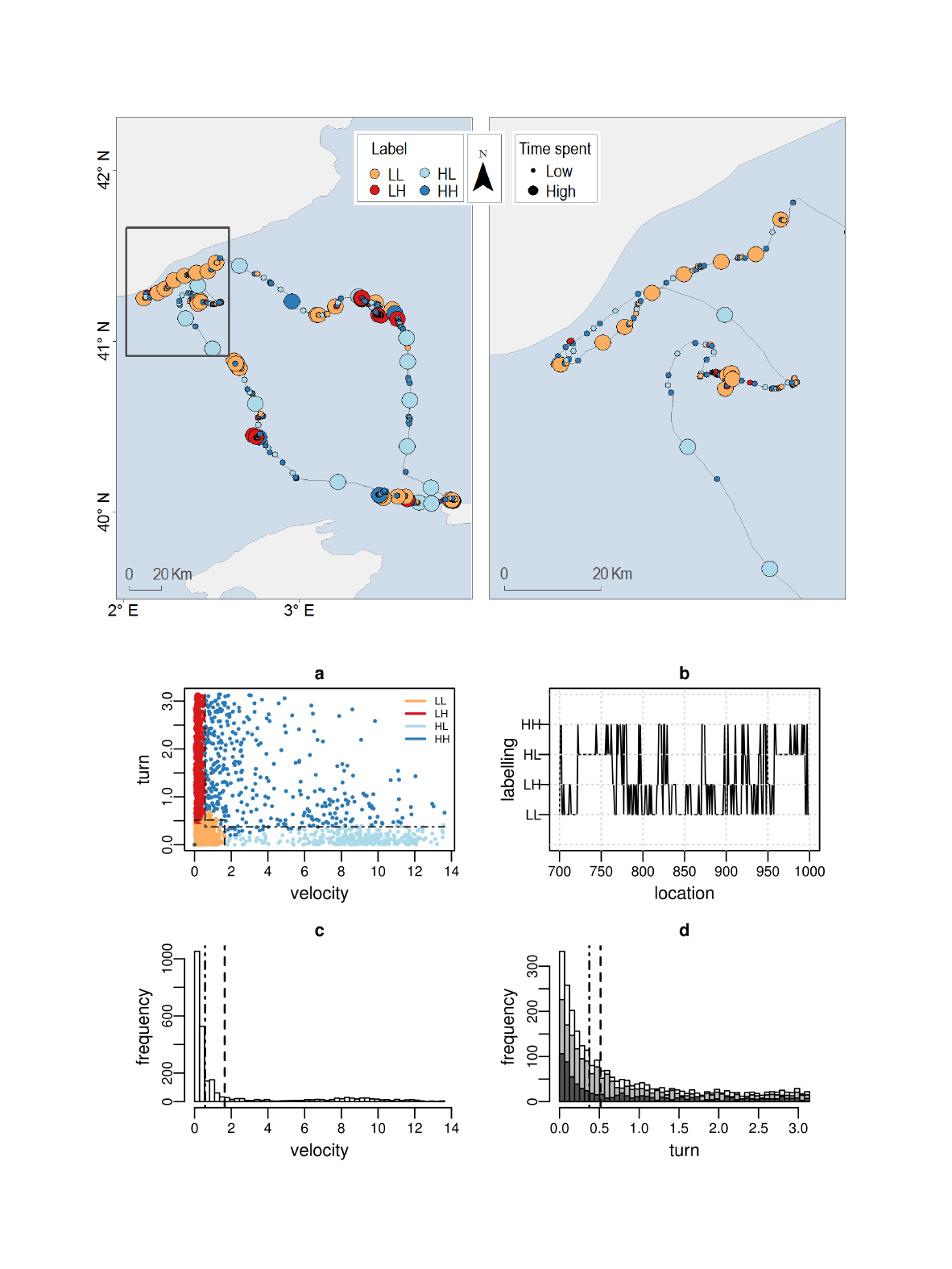}
\caption{\textbf{Cory's shearwater (\textit{Calonectris diomedea}) foraging trajectory.} Upper panel: Burst-wise labelled foraging trajectory. Symbolization by label (LL:orange, LH:red, HL:cyan, HH:blue) and time spent (symbol size: 2 natural jenks). Scale bar is only approximate. Due to copyright restrictions the figure is for representative purposes only. Source: Made with Natural Earth; Free vector and raster map data @~naturalearthdata.com. Bottom panels: (a) velocity/turn scatter plot (clustering colour code: LL:orange, LH:red, HL:cyan, HH:blue), (b) temporal behavioural profile (from location 700 to 1000) (c) velocity ($m/s$) and (d) turning angle ($rad$) frequency distributions (white colour). The turn distribution for low/high values of velocity is shown separately in light/dark grey colours, respectively. The black lines in panels a, c and d depict the delimiters $r_{.L},r_{L.}$ (dashed lines) and $r_{.H},r_{H.}$ (dot-dashed lines).}
\label{fig_07}
\end{figure}

\begin{figure}[!p]
\includegraphics[width=14cm,height=16cm]{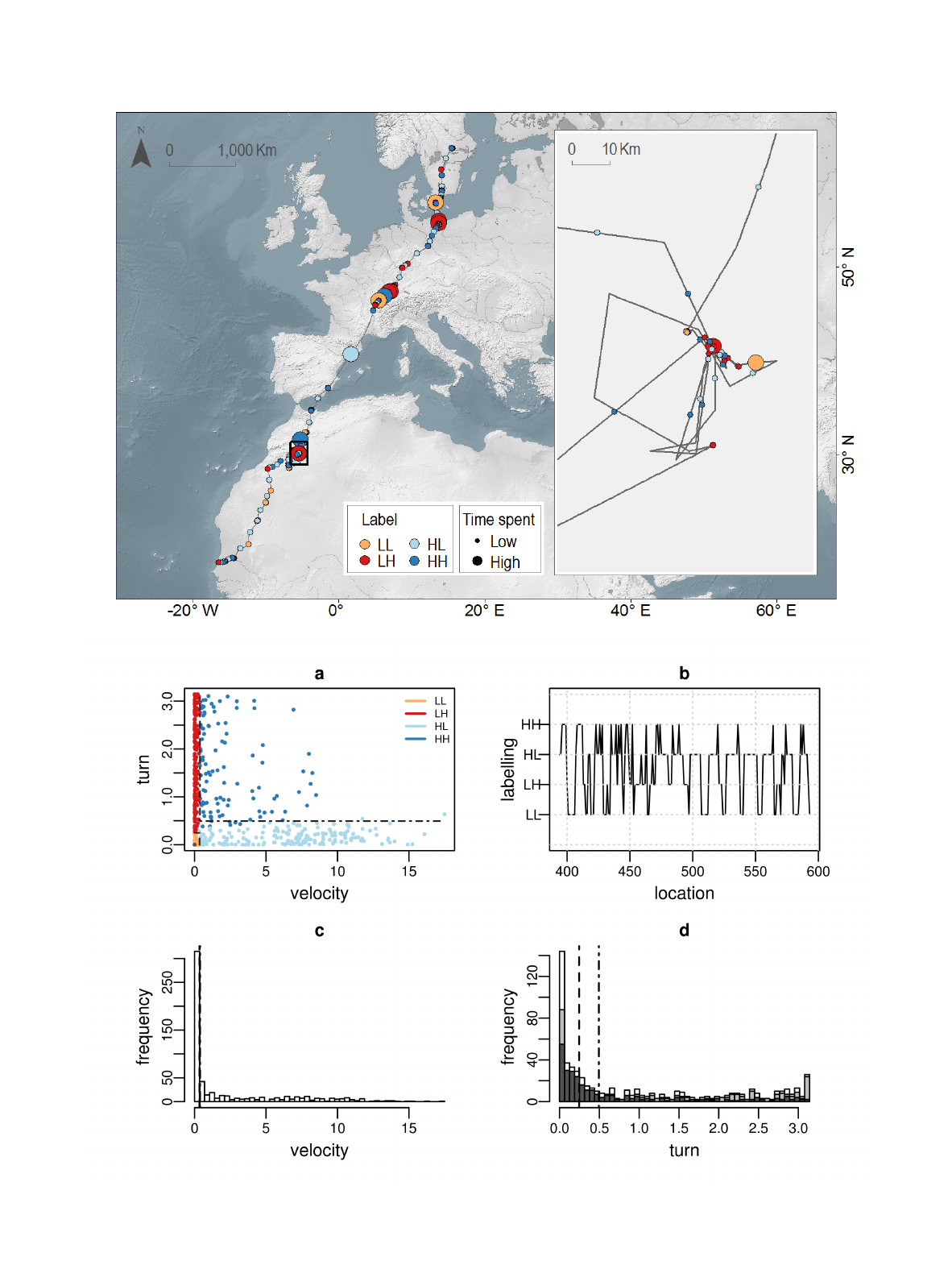}
\caption{\textbf{Osprey (\textit{Pandion haliateus}) migratory trajectory.} Upper panel: Burst-wise labelled migration trajectory. Symbolization by label (LL:orange, LH:red, HL:cyan, HH:blue) and time spent (symbol size: 2 natural jenks). Scale bar is only approximate. Due to copyright restrictions the figure is for representative purposes only. Source: Made with Natural Earth; Free vector and raster map data @~naturalearthdata.com. Bottom Panels: (a) velocity/turn scatter plot (clustering colour code: LL:orange, LH:red, HL:cyan, HH:blue), (b) temporal behavioural profile (from location 400 to 593) (c) velocity ($m/s$) and (d) turning angle ($rad$) frequency distributions (white colour). The turn distribution for low/high values of velocity is shown separately in light/dark grey colours, respectively. The black lines in panels a, c and d depict the delimiters $r_{.L},r_{L.}$ (dashed lines) and $r_{.H},r_{H.}$ (dot-dashed lines).}
\label{fig_08}
\end{figure}

The Cory's shearwater (\textit{Calonectris diomedea}, Fig.~\ref{fig_07}) and the Osprey (\textit{Pandion haliateus}, Fig.~\ref{fig_08}) datasets show a clustering layout with similar velocity/turn partition and similar semantic labelling, regardless of the ecological context (\textit{i.e.} migration, foraging) or the sampling resolution (\textit{i.e.} Argos, high-resolution-GPS), although with different proportions in the LL (resting) and HL (relocation) modes according to the ecological context (\textit{i.e.} $LL=37\%,\,HL=13\%$ for the Cory's shearwater versus $LL=18\%,\,HL=30\%$ for the Osprey, see further details in \mynameref{S2 Text}). In both, the velocity distribution (Figures~\ref{fig_07},~\ref{fig_08} panel c) shows bi-modality to some extent (although hardly apparent in Fig.~\ref{fig_08} because of the relative high frequency of low values), thus being the binary partition assumption particularly suitable. Within these standard layouts, the HH (dark blue) labelling is usually subject to more subtle
semantic interpretation. In Fig.~\ref{fig_07}a, the distribution of HH in the scatter plot suggests the existence of two possible sub-modes, one more closely related to foraging (low velocity and wide turn range) and the other more closely related to relocation (high velocity and low turns). A partition with only three clusters (LL,LH, and HL), with the HH cluster absorbed partly by the LH and partly by the HL clusters, would probably represent a better behavioural classification. However, the likelihood pay-off of this solution prevents the algorithm to reach it. Conversely, Fig.~\ref{fig_08}a shows an homogeneous HH cluster. A visual check of these data points on the landscape map reveals that they correspond to long relocations within stopover areas, thus justifying their assignment to a different behaviour.

\begin{center}
\textbf{Figure~\ref{fig_07}. Cory's shearwater (\textit{Calonectris diomedea}) foraging trajectory.}
\end{center}

\begin{center}
\textbf{Figure~\ref{fig_08}. Osprey (\textit{Pandion haliateus}) migratory trajectory.}
\end{center}

\begin{figure}[!p]
\includegraphics[width=14cm,height=16cm]{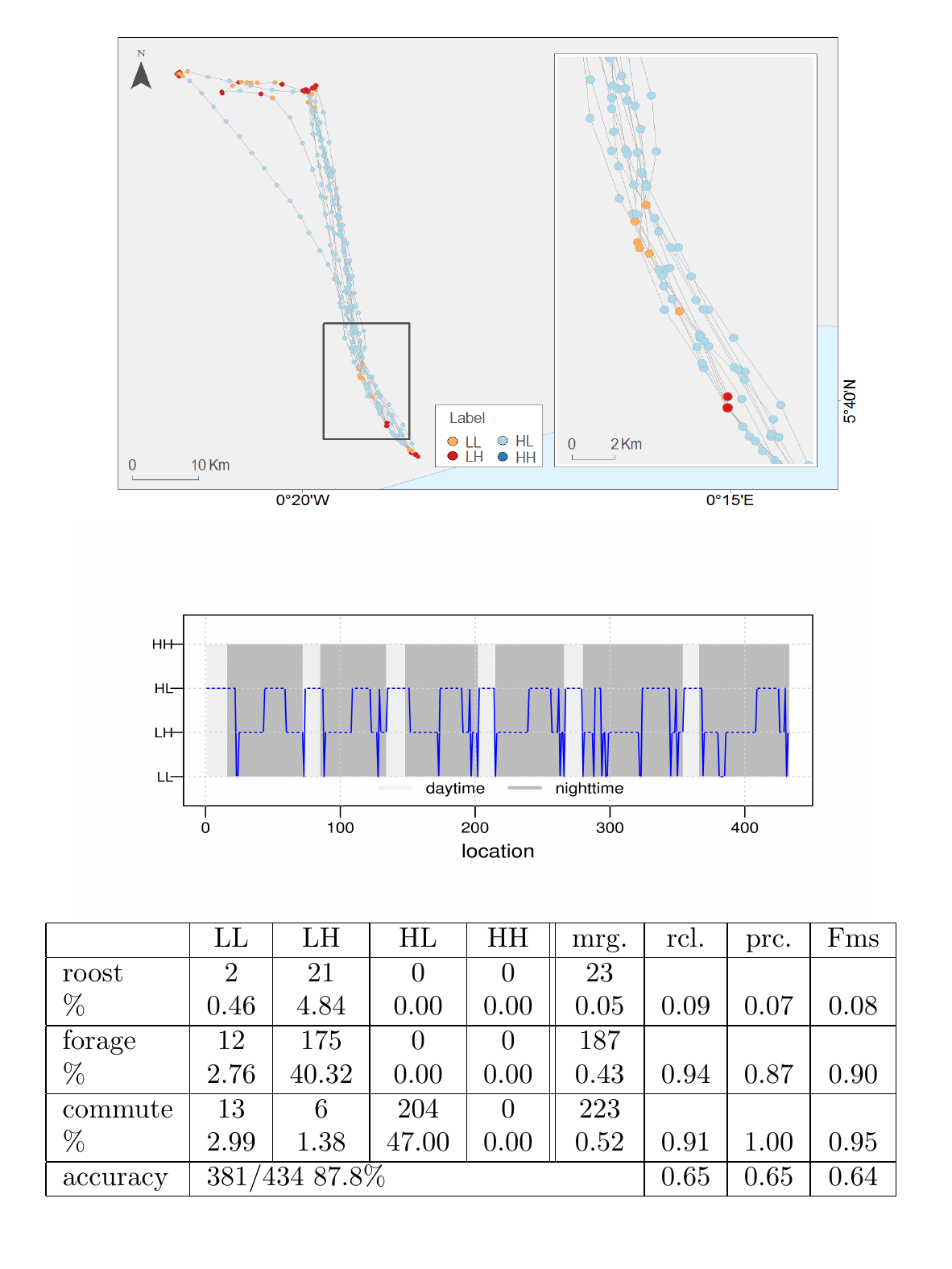}
\caption{\textbf{Bat (\textit{Eidolon helvum}) foraging trajectory.} Upper: Point-wise labelled foraging trajectory. Symbolization by label (LL:orange, LH:red, HL:cyan, HH:blue) and time spent (symbol size: 2 natural jenks). Scale bar is only approximate. Due to copyright restrictions the figure is for representative purposes only. Source: Made with Natural Earth; Free vector and raster map data @~naturalearthdata.com. Centre: smoothed temporal behavioural profile with daytime/night-time (light/dark grey) background indication. Bottom: Confusion pie showing expert \textit{vs.} EMbC labelling. Column titles \textit{mrg.}, \textit{rcl.}, \textit{prc.} and \textit{Fms} stand for \textit{marginals}, \textit{recall}, \textit{precision} and \textit{F-measure} respectively.}
\label{fig_09}
\end{figure}

The Straw-coloured fruit bat roosts in the colony during the day and moves for foraging in a very directed manner to individual fruiting trees during the night (Fig.~\ref{fig_09}). The GPS was turned off during the day and fixes occurred when the animal moved during the night. In this example we used the post-processing smoothing procedure. The behavioural labelling profile (Fig.~\ref{fig_09}, central panel) shows a quite regular behavioural pattern. It is worth noting that after the smoothing procedure, some LL labels still remain suggesting the existence of a real but short transient state (LL) occurring between HL and LH state. A few more LL labels appear also in between relocation periods (as shown in Fig.~\ref{fig_09} upper panel inset). These locations seem to correspond to specific landmarks in the daily relocations of the animal and might have some biological relevance. Indeed, it is characteristic of the EMbC algorithm to capture behaviours showing strong correlations among movement variables
despite being short in time and happening only intermittently. The decision on whether to consider or else to smooth out these type of behaviours relies on the expert decision.

For this trajectory we compared our results with an expert's labelling identifying behavioural modes (\textit{i.e. roosting, forage, commuting}) stemming from GPS and acceleration data. From the visualization of the annotated trajectory (Fig.~\ref{fig_09}, top panel) we can easily assimilate the \textit{forage} mode with the LH cluster and the \textit{commuting} mode with the HL cluster and we can therefore build the confusion matrix shown in Fig.~\ref{fig_09} bottom table. The expert classification embeds reasonably well into the EMbC classification. However, it is clear that \textit{roosting} behaviour is not well defined in terms of velocity and turn.

\begin{center}
\textbf{Figure~\ref{fig_09}. Bat (\textit{Eidolon helvum}) foraging trajectory.}
\end{center}

\begin{figure}[!p]
\includegraphics[width=15cm,height=17cm]{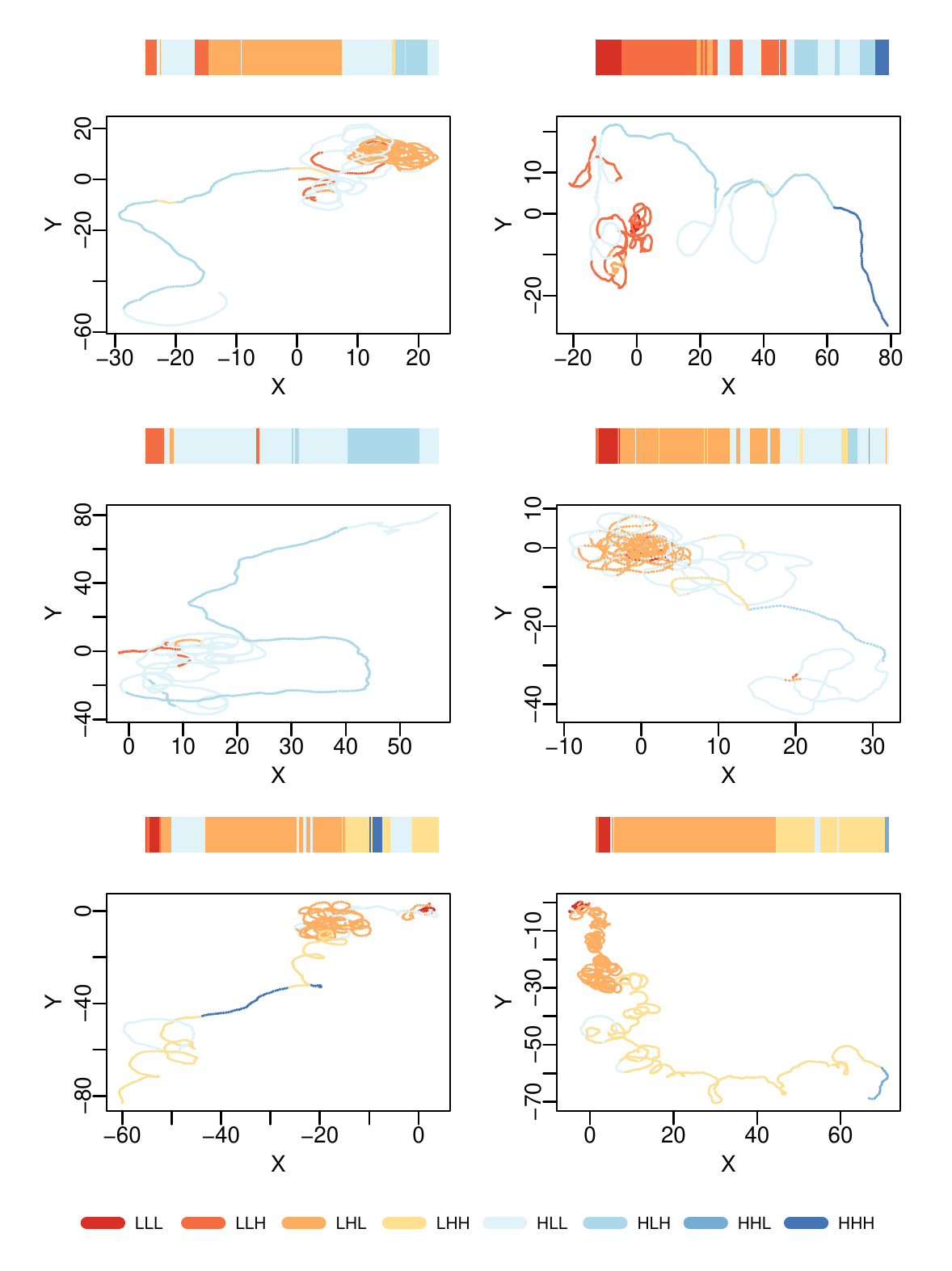}
\caption{\textbf{C.elegans (\textit{Caenorhabditis elegans}) search trajectory.} Trivariate clustering of 6 solitary nematode trajectories crawling in an agar plate (with a sampling frequency of 32 Hz and 90 minutes of trajectory time). The clustering is performed at population level (all data points at the same time) and is afterwards visualized on each individual trajectory. The algorithm captures different behaviours in terms of intensity of local search, looping behaviour and relocation.}
\label{fig_10}
\end{figure}

As an example of a trivariate clustering with different input features, Fig.~\ref{fig_10} shows a subset of results obtained when applying the EMbC algorithm at the population-level on solitary nematode crawling in an agar plate. Tracks are highly resolved ($32$Hz) and last for about 90 minutes. We computed three movement variables combining information about the shape of the trajectory and the speed of the individual (i.e. average straightness, average velocity, and net displacement) over 5 minute windows (Section~2 in \mynameref{S1 Text}). With 3 variables, the number of potential clusters is $2^3=8$. Because the number of clusters is limited, the larger the pool of individual trajectories, the more the clustering will tend to favour generic behaviours to the detriment of individual specific behaviours. In addition, only a subset of the population-level clusters are recovered in each individual, unveiling the presence of individual-level behavioural variability (Fig.~\ref{fig_10}). Accordingly to our input
features, the semantic labelling of the output clusters must be considered in terms of looping behaviour or intensity of local search and the will of the individual to move to a different location. These semantics are further explained by the statistics of the clustering given in Table~1 in \mynameref{S1 Text} Section~2.

\begin{center}
\textbf{Figure~\ref{fig_10}. C.elegans (\textit{Caenorhabditis elegans}) search trajectory.}
\end{center}

\subsection{Robustness to data loss and data inaccuracy}

\begin{figure}[!p]
\includegraphics[width=15cm,height=6cm]{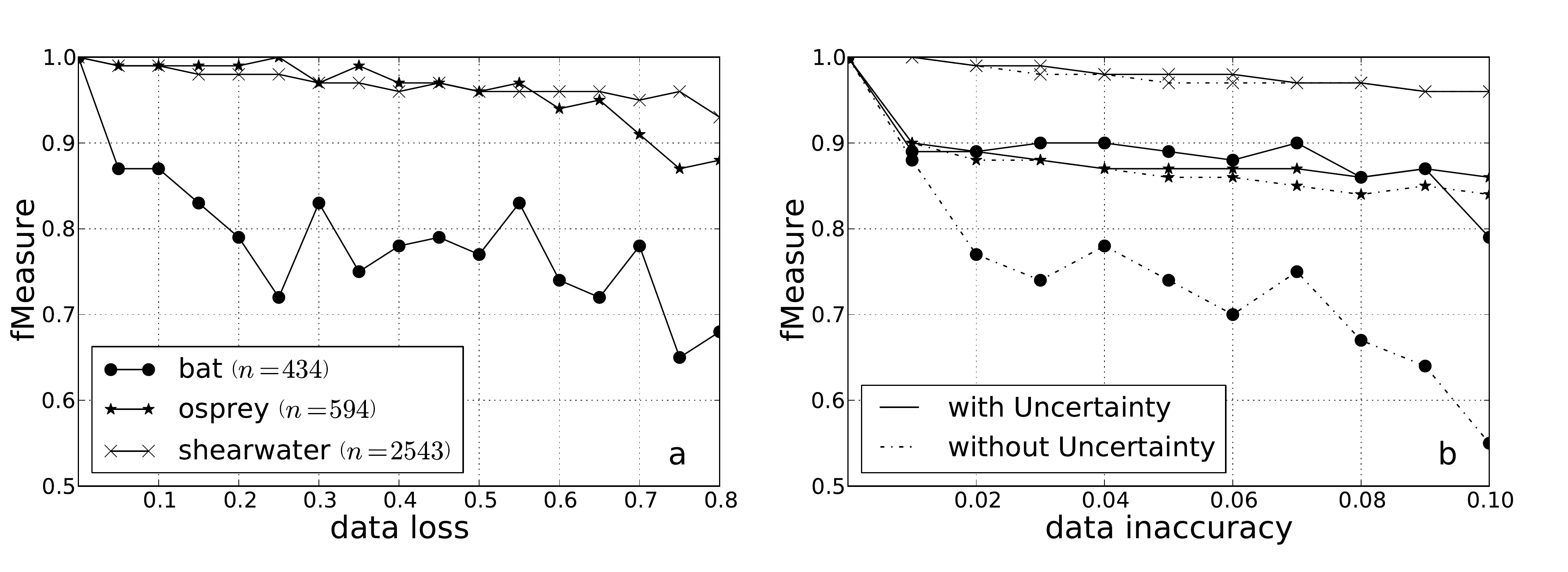}
\caption{\textbf{Data loss and data accuracy.} EMbC robustness results for a data loss range of $0\leq k_{dl}\leq 0.8$ and for a jittering range of $0\leq k_{di}\leq 0.1$, (see Equation~\ref{eq:jitteringNoise}). For each trajectory the values shown correspond to the average F-measure after 10 different runs.}
\label{fig_11}
\end{figure}

We assessed the robustness to data loss of the EMbC algorithm. In general, the larger the dataset, the more robust is the EMbC labelling to data loss (Fig.~\ref{fig_11}~a). However, the absence or presence of strong heterogeneities in the marginal distribution of clusters also plays a role. For example, although the Osprey and the Straw-coloured fruit bat datasets are both small ($n=594$ and $n=434$, respectively) the former is more robust to data loss. Interestingly, Osprey data shows more uniform posterior marginal distribution of clusters ($LL=17.51\%,\,LH=35.02\%,\,HL=29.97\%,\,HH=17.34\%$) than the bat data ($LL=10.60\%,\,LH=43.09\%,\,HL=46.08\%,\,HH=0.00\%$). As it is a nocturnal bat, the daily resting in the roost was intentionally skipped by a pre-fixed sampling scheme, and therefore, the LL cluster commonly associated to resting behaviour is misrepresented ($LL=10.60\%$). Accordingly, neither the position of the LL cluster nor its mean velocity of 2.26 m/s (see the scatter plot and statistics in \mynameref{S2 Text}) suggest such type of semantics. In general, pre-assigned GPS fixes scheduling will bias the sampling distribution of behaviours, thus conditioning both the labelling outcome and the robustness of the results to data loss.

Fig.~\ref{fig_11}~b shows the robustness of the EMbC algorithm to data inaccuracy when weighting or not the contribution to the clustering of each data point on the basis of a reliability function (Equation~\ref{eq:certainty}). In high resolution GPS datasets (\textit{e.g.} Cory's shearwater), the algorithm shows a strong robustness because inaccuracies due to sampling heterogeneity are expected to be low, so that the effect of accounting for data reliability is almost non-significant. However, accounting for data reliability in datasets with large sampling heterogeneities (i.e. ARGOS Osprey dataset) or prefixed sampling schedules (e.g. GPS Straw-coloured fruit bat dataset) favours the robustness of the labelling to data inaccuracy.

\begin{center}
\textbf{Figure~\ref{fig_11}. Data loss and data accuracy.}
\end{center}

\section{Discussion}

Splitting a trajectory into its most basic components is essential for studying and understanding mechanisms of movement \cite{Nathan2008, Fronhofer2012, Shamoun-Baranes2012}. Currently, trajectory segmentation algorithms constitute an essential component of ecological spatio-temporal analyses that seek to mechanistically understand organisms' interactions with the environment.

Current behavioural annotation methods show gradients of complexity, supervision requirements, and sensitivity to initial conditions, as well as to sampling rate and data accuracy~\cite{Jackson2011, Dean2013, Freeman2013, Joo2013, Charles2014, Bestley2010, Gloaguen2015, Metzner2015}. A possible classification of methods could result from the kind of underlying assumptions: (i) assumptions about the input feature distributions (EMC family), (ii) assumptions about time-dependency and context-dependency among behavioural states (HMM family)~\cite{Joo2013, Charles2014, Bestley2010}, and (iii) assumptions about the autocorrelation structure of the movement variable~\cite{Gloaguen2015, Metzner2015}. It is hard to make fair comparative analyses across all these methodologies in terms of computational costs, sensitivity and robustness with given datasets. In this context, we are essentially concerned about utility and simplicity of the models in the line of reasoning that \textit{all models are wrong, but some are
useful}~\cite{Box1976, Box1978}. In our case, the idea of usefulness stems from the fact that the most elementary partition of the input space (i.e. a binary partition into High/Low values of the variables) can be very informative, in many circumstances sufficient, to characterize behavioural patterns. The EMbC algorithm finds a solution for this binary partition based on a likelihood criterion under the assumption of a multivariate Gaussian mixture space. Such a key and simple concept helps reach a compromise between interpretable behavioural annotation and adequate statistical performance.

In terms of implementation, the EMbC algorithm implies iteratively computing the centroids of the clusters within regions determined by explicit \textit{delimiters} (Equation~\ref{eq:boundedMaximization2}) in order to provide easy and interpretable semantics (i.e. LL, HL, LH, HH). Nonetheless, at each iteration, we keep the computation of cluster covariances unbounded to incorporate information about the correlation landscape provided by the whole variable space (Equation~\ref{eq:boundedMaximization3}). The choice of a binary partition of variables into H/L clusters, restricts the maximal number of clusters to $2^m$, where $m$ is the number of input variables used. This restriction over the number of clusters indeed avoids \textit {aprioristic} decisions on the number of clusters in the n-dimensional space, and facilitates their interpretation.

The initial assumptions implemented in the EMbC algorithm aim at minimizing biases and sensitivity to initial conditions: (i) each data point is assigned a uniform probability of belonging to each cluster, (ii) the prior mixture distribution is uniform (each cluster starts with the same number of data points), (iii) the starting partition, (\textit{i.e.} the starting delimiters position), is selected based on a global maximum entropy criterion, thus conveying the minimum information possible. A single parameter $\sigma_{min}$ controls the minimal resolution at which clusters will be accepted. In practical terms, $\sigma_{min}$ can be directly related to measurement errors (or maximum resolution) of the variables and will limit the minimum range of variability (\textit{i.e.} minimum standard deviation) within the clusters obtained. Based on intuitive physical and biological considerations, we set $\sigma_{min}=0.01$ m/s and $\sigma_{min}=0.087$ rad (5 degrees), respectively for velocity and turn.

The algorithm deals very intuitively with data reliability: the larger the uncertainty associated with the values, the smaller the leverage of those values in the clustering. We considered two elementary sources of uncertainty: sampling heterogeneity (for those variables whose reliability depends on the sampling interval), and the measurement error of the devices. In the present work we computed velocity and turn based on the sampling intervals. Although it is worth incorporating the measurement error of the devices, here, for the sake of simplicity we considered that the major source of error comes from sampling interval lengths. However, the algorithm is multivariate and can deal with any type of movement/behavioural variables and error sources. For example, one could also use instantaneous speed~\cite{Kamran2013} or tri-axial accelerometer data~\cite{Nathan2014, Bom2014} and take into consideration the errors associated to them.

Without intending to be exhaustive, we have presented a comparison of the EMbC with similar state-of-the-art algorithms like the EMC and basic HMMs, in terms of their performance in clustering synthetically generated datasets based on GMMs. Being simpler, the EMbC yields a similar degree of performance without the need of multiple restarts or initial parametrization. Of note, HMMs are more complex in that behavioural states are defined based on both the states' definitions (the distribution of input features) and the transitions among them (the Markov-chains). Compared to HMM and EMC, we have shown that the EMbC proves particularly useful as long as: (i) we can expect bi-modality, to some extent, in the distribution of the input features, (ii) we can expect that forcing a binary partition of the input space can provide useful information, and (iii) we cannot guarantee that the temporal state dynamics can be associated to a Markov-chain process. A basic HMM is equivalent to assuming that, for an individual in
a particular state, the probability of changing to a different state remains constant as it keeps moving. In terms of movement data this is almost equivalent to assuming a memoryless individual with stationary internal states. Additionally, in movement trajectories the first-order dependence condition of a Markov-chain is easily violated because of the heterogeneity in empirical time series due to large gaps, or prefixed sampling scheduling. To overcome this problem, either we use more complex HMM approaches taking into account sampling heterogeneities~\cite{Joo2013} and/or introducing explicit time or spatial dependencies among states~\cite{Bestley2010, Gloaguen2015}, or else, we disregard any assumption about state dependence on time (EMbC). The main message, however, is that regardless of their complexity, HMM approaches are always forced to make estimates on state transitions. When the assumptions related to these transitions are not fulfilled this may impair substantially their ability to correctly
identify the states. This is the reason for the low performance of HMMs in fitting trajectory states generated from a mixture-prior sampling scheme rather than from well-defined transition matrices (see Fig.~\ref{fig_05}).

The results obtained for different empirical datasets suggest that the EMbC algorithm behaves reasonably well for a wide range of tracking technologies, species, and ecological contexts (\textit{e.g.} migration, foraging). Different layouts in the scatter plots may emerge depending on the probability distribution of the input data, the underlying mixture prior distribution and the specific set of behaviours performed by the animals along the trajectories. We also show the possibility of running the algorithm at the population level (by applying the algorithm on a pooled set of trajectories from a given population) to define \textit{average modes}, that can be visualized in single trajectories.

Importantly, the degree of match between the EMbC output and a particular set of behaviours will be highly dependent on the amount of information conveyed by the input variables regarding these behaviours, but not so much on their relative proportions/durations. Indeed, it is worth mentioning that the EMbC is good at identifying behaviours that are hardly represented in the dataset. Here we focus on velocity/turn as key movement variables for trajectory behavioural annotation \textit{e.g.}~\cite{Knell2012, Gurarie2009} but it is certainly possible to choose other behavioural or physiological data (e.g. accelerometry, stereotyped turns, body postures, metabolic rates) that correlates adequately with the behavioural classification the analyst is looking for. Even though the determination of the final semantics is variable-dependent and species-specific, the method is general and robust enough to to be used across species and tracking methodologies.

Improvements on the current EMbC implementation could come from exploring other \textit{reliability} functions and their relative contribution to the final clustering. Also, one of the drawbacks of not explicitly considering the temporal correlations in the segmentation procedure (as in HMM) is that, at coarse-scales, behavioural labels may not appear as continuous as desired, and some local labels may be considered neglectable or meaningless by the expert. We suggest considering the EMbC algorithm as a fine-scale behavioural segmentation method, with optional pre/post smoothing alternatives, the latter explicitly taking into account the weights of the labels relative to each cluster and their temporal dependencies in order to generate an aggregated coarse-grained behavioural annotation of the trajectory. We also suggest the possibility of running the algorithm at the population level (by applying the algorithm on a pooled set of trajectories from a given population). This approach smooths potential
individual variability, which might be of non-interest for certain analyses. Certainly, all these alternatives (i.e. coarse-graining, population-level analysis) may have implications in the estimations of the number and durations of the behavioural states. But this kind of problem is intrinsic to any segmentation method and only the expert's criteria and the specific goals of the study can help to justify the choices taken. All in all, behaviour should be interpreted in relative and not absolute terms as it is a multidimensional and multi-scale phenomena, and the description of behaviour will inevitably be biased by the scientific goals, the observational constraints, and the methodological choices.

To ease the complex analysis of movement behaviour and segmentation to interested users we released a ready-to-use tool (the EMbC R-package) including not only the EMbC algorithm itself but also a set of side functions for straightforward analysis (clustering statistics, clustering scatter-plot, temporal behavioural profile, smoothing function) and visualization (burst and point-wise kml doc generation) of its output. The package is compatible with the Move R-package developed by the Movebank team~\cite{move}.


\newpage
\phantomsection
\addcontentsline{toc}{section}{\numberline{}Acknowledgments}
\section*{Acknowledgments}

F.B. acknowledges the Spanish Ministry of Science and Innovation (currently MINECO) (Grants: BFU2010-22337 and CGL2010-11600-E) and the Human Frontier  Science Program (Grant: RGY0084/2011). We acknowledge support of the publication  fee by the Consejo Superior de Investigaciones Cient\'ificas (CSIC) Open Access Publication Support Initiative through its Unit of Information Resources for Research (URICI).

We thank Movebank, the Segmentation Group formed at Movebank workshops (Kamran Safi, Andrea Koelzsch, Bart Kranstauer, Luca Giuggioli). We also gratefully  acknowledge Vicen\c{c} M\'endez for revising the mathematical notation.

\paragraph*{Datasets}

We thank the following researchers for providing us with datasets and expert labelled behavioural modes:

Jacob Gonz\'{a}lez-Sol\'{i}s (Departament Biologia Animal. Universitat de Barcelona, Spain) who  has kindly  provided the  Cory's shearwater data. Funding: Fundaci\'{o}n Biodiversidad.

Raymond  Klaassen  (Migration  Ecology  Group,  Department  of  Animal Ecology. Lund University, Sweden),  who has kindly provided the Osprey trajectories, along  with accurate expert  labelling. Funding: Swedish Research Council and Niels Stensen Foundation.

Dina Dechmann (Max Plank  Institute of Ornithology, MPIO, Germany) who has kindly  povided  the  Straw-coloured  fruit bat  data  along  with accurate expert labelling.  Funding:  MPIO. Tracking  data for  these animals are available at Movebank.org.

The nematode \textit{C.elegans} dataset  was generated by Mia Panlilio in the context  of the Human Frontier Science  Program project (Grant: RGY0084/2011).   PI:  W.Ryu  (Department of  Physics.   University  of Toronto,   Canada),  co-PIs:   F.Bartumeus  (ICREA   Movement  Ecology Laboratory, CEAB-CSIC),  and I.Nemenman (Department of  Physics and Biology. Emory University, Atlanta, USA).

The data sets are all included in the \mynameref{S3 Data} zip file.

\newpage
\phantomsection
\addcontentsline{toc}{section}{\numberline{}References}
\bibliography{main.bbl}

\newpage
\phantomsection
\addcontentsline{toc}{section}{\numberline{}Tables}
\section*{Tables}

\begin{table}[!htp]
\caption{\textbf{Empirical datasets.}}
\begin{tabular}{|l|l|l|r|r|r|l|}
\hline
 & & & & & & \\
Common name & Scientific name & Type & Tracks & Fixes & $\hat{\tau}$ secs. & Context \\
\hline
Osprey & \textit{Pandion haliateus} & Argos & 1 & 594 & 3600 & migration \\
Straw-coloured fruit bat & \textit{Eidolon helvum} & GPS & 1 & 434 & 299 & roosting/foraging \\
Cory's shearwater & \textit{Calonectris diomedea} & HR-GPS & 1 & 2543 & 155 & foraging \\
Nematode & \textit{Caenorhabditis elegans} & HR-video & 6 & 10203 & 3 & search \\
\hline
\end{tabular}
\begin{flushleft} Empirical datasets. Fixes indicate the total number of locations (data points) in each data set. $\hat{\tau}$ refers to the mean time interval (in seconds) between fixes.
\end{flushleft}
\label{tbl:empiricalData}
\end{table}

\newpage
\phantomsection
\addcontentsline{toc}{section}{\numberline{}Figures}
\section*{Figures}

\addtocontents{toc}{\bigskip\bigskip\noindent{\Large\bf Supporting Information\par}}

\addtocontents{toc}{\bigskip\textbf{S1 Text}\par}
\addtocontents{toc}{Movement variables\par}
\addtocontents{toc}{\textit{Caenorhabditis elegans} movement analysis\par}
\addtocontents{toc}{Data reliability functions\par}
\addtocontents{toc}{Bursted visualization of annotated trajectories\par}
\addtocontents{toc}{EMbC likelihood dynamics\par}
\addtocontents{toc}{Generation of synthetic trajectories\par}
\addtocontents{toc}{Performance measures\par}
\addtocontents{toc}{Performance comparison results\par}

\addtocontents{toc}{\bigskip\textbf{S2 Text}\par}
\addtocontents{toc}{Generating the Results with the EMbC R-package\par}

\addtocontents{toc}{\bigskip\textbf{S3 Data}\par}
\addtocontents{toc}{Data sets zip file\par}

\end{document}